\documentclass[12pt,leqno]{book}
\usepackage{amsmath,amssymb,amsfonts} 
\usepackage{graphicx}                 
\usepackage{color}                    
\usepackage{hyperref}                 
\usepackage[numbers]{natbib}
\usepackage{makeidx}

\DeclareGraphicsExtensions{.jpg}
                
\makeindex

\title{\Huge \bfseries \mbox{ }\\Music in Terms of Science}

\author{\Large James Q. Feng, Ph.D.}
\date{ }

\begin{document}
\maketitle{}\vspace{-2cm} 

\newpage
\thispagestyle{empty}
\noindent
{\small\em Copyright \copyright \  2012 by James Q. Feng.  All rights reserved.  Draft date \today }


\newpage
\thispagestyle{empty}

\mbox{  }

\addvspace{ 48 mm }

\noindent
{\Large \it To my friends who love music and are also interested in \\scientific reasoning}

\newpage
\thispagestyle{empty}

\mbox{  }

\newpage

\addcontentsline{toc}{chapter}{Contents}
\pagenumbering{roman}
\tableofcontents
\newpage
\thispagestyle{empty}
\mbox{}

\chapter*{Preface}\normalsize
  \addcontentsline{toc}{chapter}{Preface}
To many people, music is a mystery.  
It is uniquely human, because no other species produces elaborate, 
well organized sound for no particular reason. 
It has been part of every known civilization on earth.
It has become a very part of man's need to impose his will upon
the universe, to bring order out of chaos and to endow his moments
of highest awareness with enduring form and substance.
It is a form of art dealing with the organization of tones into patterns.

Art, like love, is easier to experience than define,
because it is too large, complex, and mysterious to submit to any single
adequate definition.
One of the characteristics of art is its unreasonableness.  
Many other human creations have an obvious reason 
for their necessity, usefulness, and functionality.
It seems that music serves no obvious adaptive purpose.
Charles Darwin, in ``The Descent of Man'', noted that
``neither the enjoyment nor the capacity of producing musical notes are 
faculties of the least direct use to man in reference to his
ordinary habits of life.''
Unwilling to believe that music was altogether useless,
Darwin concluded that it may have made man's ancestors more
successful at mating.  
Yet there is no evidence of one gender to be musically more gifted than
the other.  

It has been recognized that music is ingrained in our auditory, cognitive
and motor functions. 
People know much more about music than they think.
They start picking up the rules from the day they are born, 
perhaps even before, by hearing it all around them.
Very young children can tell whether a tune or harmony is right or not.
One of the joys of listening to music is a general familiarity with 
the way it is put together:
to know roughly what to expect, then to see in what particular ways
your expectations will be met or exceed.
Most adults can differentiate types of music even 
without proper training.

Despite of cultural differences, music from different civilizations seems
to consist of some building blocks that are universal:
melody, harmony, rhythm, etc.
Almost all musical systems are based on scales spanning an octave--the note
that sounds the same as the one you started off with, but at 
a higher or lower pitch.
It was discovered by Pythagoras, a Greek philosopher who lived
around 500 BC, that the note an octave higher than another has 
a frequency twice high. 
The notes that sound harmonious together have simple rational number ratios
between their frequencies.

It is those implicit structures and relationships in 
apparently mysterious musical experience that
I am interested in exploring here.
As a scientist by training with a consistent passion for classical guitar playing, I would like make an attempt 
to explain the musical experience in terms of science 
and mathematics,  hoping to fill some gaps between the knowledge of
scientists and artistic intuition of musicians.

\vspace{ 10 mm }

\noindent
James Q. Feng

\noindent
Maple Grove, Minnesota

\thispagestyle{plain}

\newpage

\pagestyle{headings}
\pagenumbering{arabic}


\chapter{Physics of Sound}
In a physical sense, sound that we hear is a mechanical (longitudinal) 
wave---the back-and-forth
oscillatory motion of matter, such as air, 
produced from a vibrating object, such as a plucked string or 
a vibrating air column in a pipe
(on a musical instrument).

\section{The vibrating string}
A stretched string, having two ends fixed on a musical instrument, tends to
maintain its straight equilibrium shape, when 
the effect of gravity is negligible, 
which is the state of the lowest energy.
When the string is pulled transversally away from its equilibrium form, 
its tension wants to bring it back to the equilibrium position.
Thus, after the string is plucked and released, it starts to move toward
its equilibrium position.
As it reaches the equilibrium position, however, it acquires certain velocity
or momentum such that it keeps moving away from the equilibrium position in
the other direction decelerating until stopped by the action of tension,
and then moves back toward the equilibrium position, and so on.
Therefore, the string vibrates back-and-forth about its equilibrium position.

\begin{figure}[h] \label{VibratingString}
\includegraphics[scale=0.64]{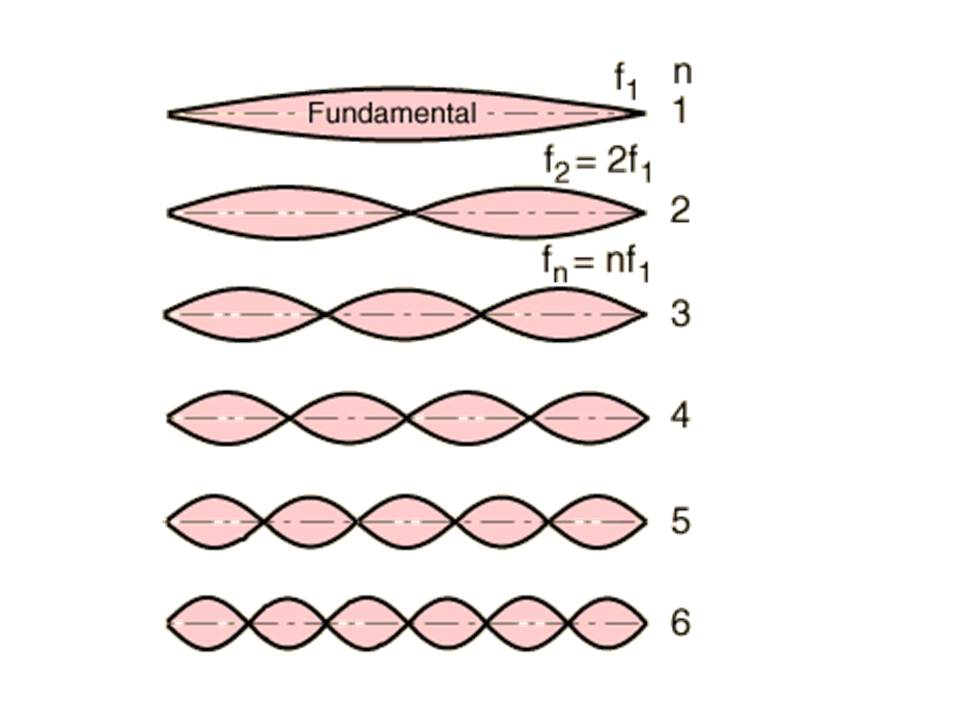}
\caption{The vibrational components of a string with two ends fixed corresponding to their frequencies
relative to that of the fundamental frequency $f_1$.}
\end{figure}

Governed by physical laws, the string under tension with two ends fixed 
can only vibrate at certain frequencies. 
The mathematical equation describing a (small-amplitude) vibrating string
(with viscous damping being neglected)
is actually in a form of the so-called (one-dimensional) ``wave equation''
\begin{equation} \label{wave_eq1d}
\frac{\partial^2 u}{\partial t^2} = 
\frac{T}{\mu} \frac{\partial^2 u}{\partial x^2} \, ,
\end{equation}
where $u$ denotes the displacement of the string from its 
equilibrium position, $t$ the time, $T$ the tension of string,
$\mu$ the mass per unit length of string (or line density),
and $x$ the coordinate (or the distance) along the string \citep{rayleigh45, landau59e}.
With the two ends of the string fixed, namely $u = 0$ at $x = 0$ and $x = L$,
the mathematical solution to (\ref{wave_eq1d}) is
\begin{equation} \label{string_vib_soltn}
u = 
\sum_{n=1}^\infty A_n \cos(\omega_n t + \alpha_n) 
\sin \left(\frac{n \pi x}{L}\right) \, ,
\end{equation}
where $\omega_n \equiv (n \pi/L) \sqrt{T/\mu}$ denotes the angular frequency
with $A_n$ and $\alpha_n$ being determined from the initial 
deformation of the string from its equilibrium form 
before the string is released.
Equation (\ref{string_vib_soltn}) indicates that
a string (of uniform line density) under (constant) tension with
two fixed ends vibrates as superposition of 
a series of ``standing waves''\footnote{A standing wave
is a wave that remains in a stationary position,
as a result of interference between two waves 
(of the same frequency and amplitude) traveling in 
opposite directions} \index{Standing wave}
each with an angular frequency ($\omega_n = 2 \pi f_n = 2 \pi n f_1$ 
where $f_n$ is the normally called frequency) and a wavenumber given by 
an integer $n$ multiplying the ``fundamental frequency'' $\omega_1 = (\pi/L)\sqrt{T/\mu}$
and the ``fundamental wavenumber'' $\pi/L$ 
(corresponding to the ``fundamental wavelength'' $2 L$). \index{Fundamental frequency}
In terms of musical tuning, 
the fundamental frequency defines the ``pitch'' of the ``tone'' and 
is used to name the note,
whereas those higher frequencies are called ``overtones'' 
that influence the ``tone color''. \index{Overtones}
Thus, the pitch of a vibrating string 
can be adjusted by changing the tension $T$ (e.g., by turning the tuning peg),
the line density $\mu$ (which is often related to the thickness of the string),
and the length between the two fixed ends 
(which is determined by the position on the instrument fingerboard where 
the string is pressed by a finger).

\section{The sound wave}
When the string is vibrating, the sounding board connected to it is also excited to vibrate at the same frequency.
With its large vibrating area, the sounding board effectively drives the neighborhood air molecules into oscillatory motion,
which produces the oscillatory disturbance in the air.
Such disturbance propagates in the form of a (longitudinal or compressional) mechanical wave---the sound wave,
according to the physical laws.
In general, the sound wave propagation (in an ideal fluid) can be described by
the wave equation
\citep{rayleigh45, landau59f}
\begin{equation} \label{wave_eq}
\frac{\partial^2 \phi}{\partial t^2} = 
c^2 \nabla^2 \phi \, ,
\end{equation}
where $\phi$ denotes the velocity potential, 
$\nabla^2$ the Laplacian operator, 
and $c \equiv \sqrt{(\partial p/\partial \rho)_S}$ 
the speed of sound (which relates to the adiabatic or isentropic  
compressibility of the fluid
(with $p$, $\rho$, $S$ denoting the pressure, density, and entropy,
respectively).
Under standard atmospheric condition, the speed of sound in air 
is about $340$ meters per second. \index{Speed of sound (in air)}
Within a wide range, the speed of sound in air appears to be 
independent of the frequency and intensity of the sound.
Otherwise, a piece of music would be heard at a distance hopelessly
confused and discordant \cite{rayleigh45}.

For a given frequency $\omega$, (\ref{wave_eq}) can have 
a solution of the traveling wave form
\begin{equation} \label{wave_eq_soltn}
\phi = A \, \exp[i(\omega t - \mbox{$\bf k$} \cdot \mbox{$\bf r$})] \, ,
\end{equation}
where $\mbox{$\bf k$}$ is the wave vector 
and $\mbox{$\bf r$}$ the position vector. 
Because the wavenumber or the magnitude of the wave vector 
$|\mbox{$\bf k$}| = 2 \pi/\lambda$ is equal to $\omega/c$,
the wavelength $\lambda$ for a $30$ hertz sound wave \index{Wavelength of sound}
is about 11 meters, and $~1.1$ meters for a $300$ hertz sound wave,
$~0.11$ meters for a $3000$ hertz sound wave, ...
If an object is in the path of propagation of a sound wave
and the wavelength of which is large compared with the dimension of 
the object,
scattering of sound can be significant with the scattering cross-section 
proportional to $\omega^4$ \citep{landau59f}.
On the other hand, the concept of sound ray 
(similar to that in geometric optics) is usually applicable 
for a sound wave of wavelength much smaller than
the dimension of the object.
In reality, the sound energy is progressively dissipated due to 
the existence of viscosity and thermal conductivity in fluid. \index{Sound dissipation}
It can be theoretically shown that the sound intensity 
decreases with the traversed distance $x$ according to a law 
$\exp(-2 \gamma x)$ with the ``absorption coefficient'' 
$\gamma \propto \omega^2$ \citep{landau59f}.
Hence, the high frequency components of sound decays much faster with
propagation distance than the low frequency components.

\section{Hearing the sound}
When the sound, as a series of alternating compressions and rarefactions
of air, propagates and reaches our ears,
the variation of pressure in the air is transferred  
via external ear and ear canal to the ear drum, ear bones, 
and cochlea in the inner ear where the vibration signal is processed.
Our eardrum is simply a membrane that is stretched across tissue and bone.
It serves as the gate of our auditory perception through which we sense 
what is out there in the auditory world.
Sound is transmitted through the air by molecules vibrating at certain frequencies and amplitudes.
In other words, 
what is sensed by our ears is a time-varying (pressure) signal 
(due to air molecules bombarding the eardrum)
that can be expressed as a function of time or a function of frequency, 
e.g., a Fourier integral for a somewhat noisy sound or a Fourier series for a musical sound like that
produced by a vibrating string as given by (\ref{string_vib_soltn}).\footnote{Actually (\ref{string_vib_soltn}) 
can be considered as a general mathematical expression of a Fourier series for a musical sound perceived by our ears}
\index{Fourier series}

\begin{figure}[h] \label{WaveFormsOfDiffInstr}
\includegraphics[scale=0.64]{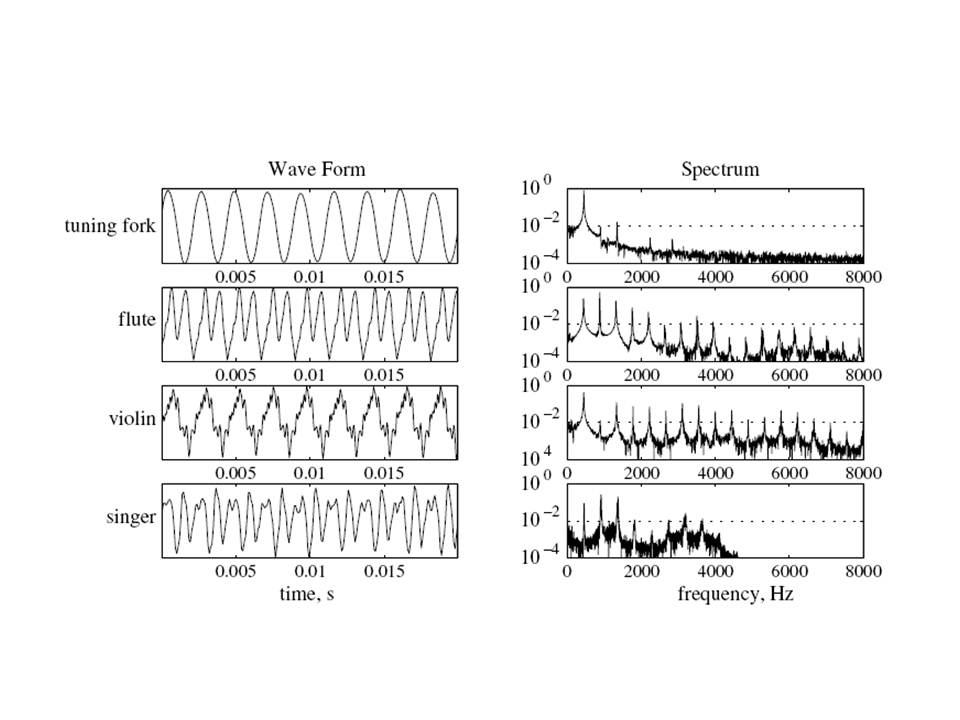}
\caption{Acoustic pressure waveforms and their frequency spectra of note A (at 440 Hz) played on different instruments.}
\end{figure}

In fact, the same musical note played on different instruments sound differently 
due to different vibration mechanisms of the sound source.  
A musical note is defined by its fundamental frequency only; for example, the note A4 has a frequency of 440 Hz.
But when played on any musical instruments, the note A4 
almost always contains many higher overtone harmonics in addition
to that of its fundamental frequency of 440 Hz (see figure 1.2). 
It is the relative values of the coefficients $A_n$ in (\ref{string_vib_soltn}) 
that determines the quality or color of the note we hear,
though oddly its phase $\alpha_n$ does not have an effect on our perception of the sound 
despite its obvious effect on changing the waveform. 
What a musical note describes is the pitch that we perceive. 
In general, pitch is a subjective sensation, 
which can be comprehensively fooled to create auditory illusions. 
Yet the perceived pitch is nearly always closely connected with 
the fundamental frequency of a note,
with lesser connection to sound intensity, harmonic content of the sound, 
and to the immediately preceding history of notes heard.
However, even for tones of equal intensity, 
the perceived pitch and measured frequency may not stand in 
a simple linear relationship.

As a wave, the sound that is received by our ears may be distorted 
from that emitted right from the source during its propagation in air, 
e.g., by the Doppler effect due to air flow, 
diffraction, interference, etc.,
because the vibrating molecules at our eardrum tell nothing about 
the source of the vibration and the path of sound wave propagation. 
Thus, the well balanced music produced by the musicians on the stage 
may not be heard as much so by the audience at the back seats
just because of the frequency dependent dissipation of sound energy in air
(if nothing else). 
The lower notes in a music would be heard relatively louder 
at a distance because 
they decay slower than the higher notes.

But in hearing music, one may be surprised by the fact that the entirety of 
a piece of music is often not compromised by some of the missing or altered physical elements 
such as a missing note or a wrong note by mistake or 
occasionally distorted sound wave forms.
More often than not, our brains are able to ``fill in'' the missing information or to ``correct'' the 
``wrong note'' based on expectations according to certain rules for
harmonic progression and cadence.  
Many great musicians often made little technical mistakes like 
a wrong note, a rushed note, etc. while their performances were filled with passion 
and can touch the heart and soul of their audience
(which are what actually made them great).
In contrast, some technical wizards may be capable of rendering a flawless 
performance and playing all the notes correctly (in the strict, 
as-notated sense), but they may fail conveying the true meaning.
In reality, what most of people turn to music for is an emotional experience.\footnote{A
sad fact is that most of the music school trainings have looked for 
accomplishment in the facility of fingers rather than the expressiveness of emotion.
It is also a fact that how to effectively communicate with or 
move a listener is much more difficult to teach, 
if not totally impossible.}

\newpage 
\thispagestyle{empty}


\chapter{Musical Tuning}

One of the basic elements in music is the melody 
that is organized rhythmically 
with a discrete set of tones or pitches called the musical scale
to express a musical idea.
(Here a tone is a discrete musical sound, whereas a pitch, as a purely psychological construct,
relates both to the actual frequency of a particular tone 
and to its relative place in the musical scale.)
The relationship among those tones or pitches exhibits 
a remarkable array of number properties.
Another important musical element is the harmony that is built upon
chords with a number of tones,
which brings out the sense of consonance and dissonance, 
the feeling of tension and harmonic motion, etc. 
according to the frequency ratios of those tones.

\section{The musical scale}
Among many possibilities, the most important scale
is the diatonic scale, \index{Heptatonic scale}
which is a seven-note or seven-tone (heptatonic), octave-repeating musical scale
comprising five whole steps and two half steps in each octave. 
The two half steps are separated from each other 
by either two or three whole steps,
to ensure that, in a diatonic scale spanning more than one octave, 
all half steps are maximally separated from each other.\cite{burns99} \index{Musical scale}

Although a simple definition of musical scale is difficult to give,
an example is fairly simple to find, such as the best known major scale:

\begin{center}
\begin{tabular*}{0.75\textwidth}{@{\extracolsep{\fill}} lccccccc }
  do & re & mi & fa & so & la & ti & do  \\
\end{tabular*}
\end{center}

\noindent
which virtually everyone can recite, and which is immortalized in a
song from the movie ``Sound of Music''.
For the C-major scale, ``do'', ``re'', ``mi'', ``fa'', ``so'', ``la'', 
``ti'', and ``do'' correspond to the white keys on a piano keyboard
``C'', ``D'', ``E'', ``F'', ``G'', ``A'', ``B'', and ``C''.
Each tone or pitch is referred to as a ``degree'' of the scale. \index{Degree of scale}
The distance between two successive tones in a scale is called a scale step.  \index{Scale step}
The tones or notes of a scale are numbered by their steps from the root of the scale (the ``tonic''). 
(Here the words tone and note both refer to the same entity in the abstract,
where the word ``tone'' refers to what you hear, and the word ``note'' 
to what you see written on a musical score.)
For example, in a C-major scale the first note is C (the 1st-degree, Tonic---key note),
\index{Key note} \index{Tonic}
the second D (the 2nd-degree, Supertonic), the third E (the 3rd-degree, Mediant), 
\index{Supertonic} \index{Mediant}
the forth F (the fourth-degree, Subdominant), \index{Subdominant}
the fifth G (the fifth-degree, Dominant), \index{Dominant}
the sixth A (the sixth-degree, Submediant), and the  \index{Submediant}
seventh B (the seventh-degree, Leading tone), \index{Leading tone}
 with the eighth-degree called Tonic again or Octave. \index{Octave}  
Two notes can also be numbered in relation to each other such as
C and E (or F and A, or G and B) create an interval of a (major) third;  
D and F (or E and G, or F and A, or A and C, or B and D) create 
a (minor) third; 
C and F create a fourth, C and G a fifth, C and A a sixth, 
and C and B a seventh.

\subsection{Just intonation} \index{Just intonation}
Based on 
our perception of harmonic relationships between
frequencies, those notes with frequencies that are harmonically related
(e.g., having ratios of rational numbers with small integers)
tend to sound good together.
The ``Just Scale" (sometimes referred to as ``harmonic tuning''
or ``Helmholtz's scale'') comes out naturally as a result
of the ``overtone series'' from vibrating strings or air columns
(of musical instruments).
The octave is the fundamental interval for which the notes 
are related in frequency exactly by a ratio of $2:1$. 
With this simple frequency ratio, the notes to the ear sound 
as if they are ``the same note''.\footnote{In fact, when men and women
sing the same note in unison, their voices are normally an octave apart.
Children generally sing the same note an octave or tow higher than adults.}
In a C-major (just) scale, C, D, E, F, G, A, B, and C have 
the frequency ratios
(relative to C) given in Table 2.1. 
Here the perfect fifth, having a ratio $3/2$, is the most consonant interval
other than the octave $2/1$.
The next most consonant interval is the perfect fourth $4/3$, 
as the inversion of perfect fifth (i.e., 
the fifth note from C toward the lower octave
$1/2$ and then move one octave up by multiplying it by $2$,
which can be obtained as $(1 \div 3/2) \times 2$.

\begin{table} \label{FreqRatios}
\caption{The frequency ratios of just intonation}
\begin{center}
\begin{tabular*}{0.75\textwidth}{@{\extracolsep{\fill}} lccccccc }
\hline
\hline
\\
  do & re & mi & fa & so & la & ti & do  \\
\hline
\\
  C & D & E & F & G & A & B & C  \\
  $1$ & $9/8$ & $5/4$ & $4/3$ & $3/2$ & $5/3$ &
$15/8$ & $2$  \\
\hline
\hline
\end{tabular*}
\end{center}
\end{table}

It might be noted that in the just intonation,
the perfect major third has a frequency ratio $5/4$ whereas 
the perfect minor third $6/5$ ($= 3/2 \div 5/4$ or $= 2 \div 5/3$ or
$= 2 \times 9/8 \div 15/8$ but $\ne 4/3 \div 9/8$).
Thus the perfect fifth can be ideally sub-divided into a 
perfect major third and a perfect minor third such as
$3/2 = 5/4 \times 6/5$.
This just intonation is also called ``Ptolemaic tuning'', named after \index{Ptolemaic tuning}
the Greek astronomer Claudius Ptolemy (AD 85-165) who first proposed it,
to build the scale based on perfect thirds instead of 
that based on perfect fifth as attempted by Pythagoras (569-475 BC).

Most people find the harmonies in just intonation quite pleasing;
instruments tuned in this manner can really ``sing''. 
There are nevertheless a few problems.  
For example, the interval of minor third D and F 
($= 4/3 \div 9/8 = 32/27 \ne 30/25$) is not a perfect minor third.
The perfect second $9/8$ is only valid for the intervals of C and D,
F and G, A and B, but not for that of D and E, G and A, which 
are actually $10/9$.
The semitone interval between E and F, B and C are both $16/15$ though.
But $16/15 \times 16/15 \ne 9/8$ and $\ne 10/9$.  
Thus the semi-tone intervals cannot be uniformly distributed 
in an octave.

The just intonation actually belongs to a special class of 
the tuning methods called ``perfect tuning'' among which 
are also the Pythagorean tuning, mean-tone temperament, etc.\index{Pythagorean tuning}
All of those ``perfect tuning'' methods are based on 
an attempt to keep certain intervals perfect (with perfect integer ratios).
For example, Pythagoras (569-475 BC) proposed a division of the octave 
into intervals based on the ratios $2:1$ and $3:2$,
which he had discovered correspond to natural harmonies.
Thus, the interval between $1$ and $3:2$ is $3:2$ (perfect fifth), whereas 
that between $3:2$ and $2:1$ is $2 \div 3/2 = 4/3$ 
which is actually the perfect fourth. 
Then the interval between perfect fourth and perfect fifth 
is $3/2 \div 4/3$ $= 9/8$ which defines the perfect second.
If the major third is constructed by stacking a perfect second 
on top of another perfect second, it should be $9/8 \times 9/8$ $= 81/64$ 
with the semi-tone interval being $4/3 \div 81/64$ $= 256/243$.
Here we cannot build a perfect second from stacking two semi-tone intervals. 
In fact, $256^2/243^2 = 1.10986$ noticeable different from $9/8 = 1.125$.
The major third in Pythagorean tuning is $81/64 = 1.26563$ 
quite different from that in the just intonation $5/4 = 1.25$. 
  
It seems though none of those perfect tuning
methods can yield a perfect semi-tone interval to 
equally divide an octave interval. 
This problem becomes evident when we try to expand our tuning to
the chromatic keys (the black keys on a piano) and 
to transpose a song to different keys,
because of the difference in intervals between various notes. 

\subsection{Equal temperament} \index{Equal temperament}
In the modern world the universally accepted solution to
the imperfections in all ``perfect tuning'' systems, such as the
Pythagorean tuning, just intonation, mean-tone temperament,
is to distribute the unavoidable imperfections in all the notes
such that an octave is equally divided in $12$ semi-tone intervals.
This kind of twelve-note musical scale, with $12$ equally distributed 
semi-tones, is called the (equal-tempered) chromatic scale.

Thus the frequency ratio of the adjacent semi-tone notes is equal to
$2^{1/12} = 1.059463...$ whereas the perfect semi-tone interval in
the just intonation is $16/15 = 1.066666...$. 
Then all the intervals of second have the frequency ratio between the notes
as $1.059463 \times 1.059463 = 1.12246$, 
whereas $9/8 = 1.125$ and $10/9 = 1.11111...$. 
The frequency ratio of the interval of minor third is 
$1.12246 \times 1.06667 = 1.19729$ whereas $6/5 = 1.2$,
and that of major third is $1.12246^2 = 1.25992$ whereas $5/4 = 1.25$.
For the interval of fourth, the frequency ratio is
$1.12246^2 \times 1.059463 = 1.33483$ whereas $4/3 = 1.33333...$.
For the interval of fifth, it is 
$1.12246^3 \times 1.059463 = 1.49829$ whereas $3/2 = 1.5$.

An equal-tempered chromatic scale is a nondiatonic scale having no tonic,
because of the symmetry of its equally spaced tones.
Those equally spaced $12$ semi-tone notes, if all used in a melody, do not cause us to 
feel varying amounts of tension and resolution necessary for invoking a sense of motion.
Thus, a subset of seven (or less often, five) of those twelve tones is used instead 
in musical melody composition.
In a seven-note scale, a scale step consisting of two semi-tone intervals is a whole tone or whole step
whereas that of one semi-tone interval a half step.
With the equal temperament, the frequency ratios between the two notes of all the whole steps are the same,
so are that of all the half steps.
Thus, all (seven-note) scales sound similar independent of which key to start. 
The 12-tone equal temperament enables natural sounding scales when shifting keys in a composition
and transposing a song to a different key.
However,  in the equal temperament scale, 
the seventh semitone note above a note is slightly flat from the perfect fifth,
and the fifth semitone is slightly sharp from the perfect fourth, as the compromise for
obtaining a uniformly distributed frequency spacing.

\subsection{Circle of fifths} \index{Circle of fifths}
In music, besides the familiar major scale do-re-mi-fa-so-la-ti-do, there is also a minor scale
(which goes with la-ti-do-re-mi-fa-so-la).
With regard to intervals, the words major and minor just mean large and small, so a major third has a wider interval,
and a minor third a relatively narrower one.  
The intervals of the second, third, sixth, and seventh (and compound intervals based on them) may be major or minor.
A major scale is one whose third degree is a major third above the tonic, \index{Minor scale}
while a minor scale has a minor third degree. 
What distinguishes major keys from minor ones is whether the third scale degree is major or minor.
The relative minor of a major key has the same key signature and starts down a minor third 
(or equivalently up a major sixth).
For example, the relative minor of C major is A minor (which also uses only the white keys
of the piano keyboard), 
and the relative minor of G major is E minor.
This alteration in the third degree changes the mood of the music without changing the root and overall key and tonality; 
music based on minor scales tends to 
sound dark, soft, and introverted whereas that on major scales clear, open, extroverted.

\begin{figure}[h] \label{CircleOfFifths}
\includegraphics[scale=0.70]{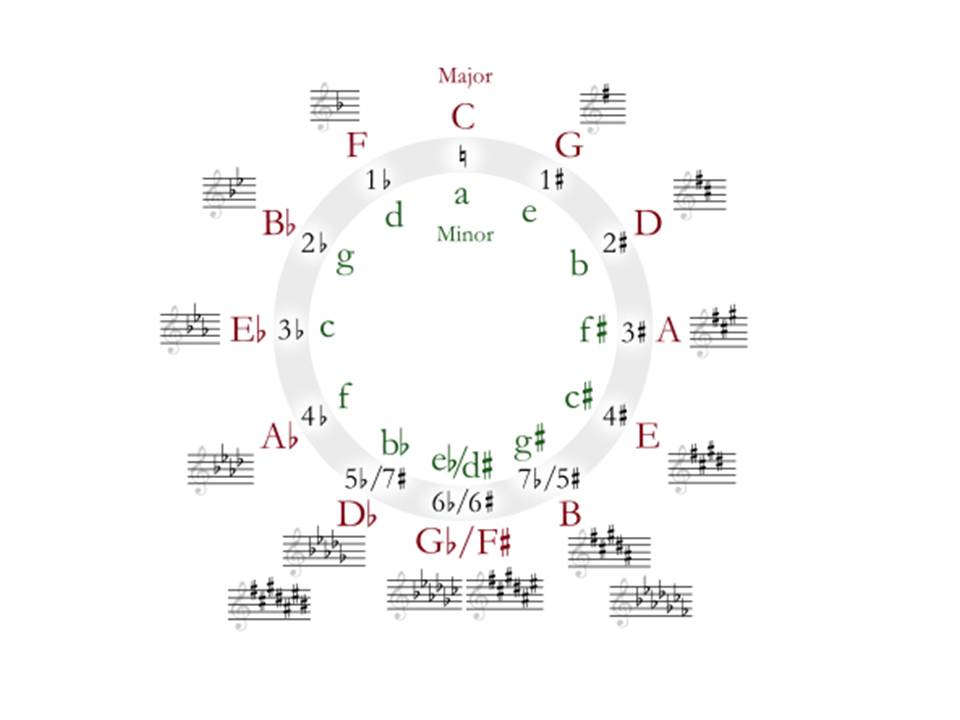}
\caption{Circle of fifths showing major and minor keys with the corresponding key signatures.}
\end{figure}

The circle of fifths (or circle of fourths) is a geometrical representation of
the relationships among the 12 tones of the chromatic scale,
their corresponding key signatures (a series of sharp or flat symbols on the staff), 
and the associated major and minor keys (as shown in figure~2.1). 
\index{Key signature}
Since the `fifth' defines an interval that is the most consonant non-octave interval
(consisting of the ratio of smallest integers $> 1$),
the circle of fifths is a circle of closely related pitches or key tonalities.
It is helpful in understanding tone relationship, in composing and harmonizing melodies,
building chords, and shifting to different keys within a composition.

At the top of the circle, the key of C Major has no sharps or flats. 
Starting from the apex and proceeding clockwise by ascending fifths, the key of G has one sharp,
the key of D has 2 sharps, and so on.
Similarly, proceeding counterclockwise from the apex by descending fifths,
the key of F has one flat, the key of B-flat has 2 flats, and so on.
At the bottom of the circle, the sharp and flat keys overlap, showing pairs of enharmonic key signatures.

Starting at any pitch, ascending by the interval of an equal tempered fifth, one passes all twelve tones clockwise,
to return to the beginning pitch class.
To pass the twelve tones counterclockwise, it is necessary to ascend by perfect fourths, rather than fifths.

\section{Beats} \index{Beats}
When two pitches of two different frequencies sound together, 
we may hear so-called beats.  
In terms of trigonometric functions, for example, 
the effect of superposition of
two pitches of two different angular frequencies $\omega_1$ and $\omega_2$
may be simply expressed as
\begin{equation} \label{beat-eq1}
u = \sin(\omega_1 t) + \sin(\omega_2 t)
\end{equation}
where we can also write
$\omega_1 = (\omega_1 + \omega_2)/2 + (\omega_1 - \omega_2)/2$
and $\omega_2 = (\omega_1 + \omega_2)/2 - (\omega_1 - \omega_2)/2$.
Thus we can also have 
\begin{equation} \label{beat-eq2}
u = 2 \cos(\Delta \omega t) \sin(\bar{\omega} t)
\end{equation}
where $\Delta \omega \equiv (\omega_1 - \omega_2)/2$ 
and $\bar{\omega} \equiv (\omega_1 + \omega_2)/2$
denoting the frequency difference and average frequency, respectively. 

When $\omega_2$ is very close to $\omega_1$,
$\Delta \omega$ is expected to be small and $\bar{\omega} \approx \omega_1$
or $\omega_2$. 
The factor $2 \cos(\Delta \omega t)$ acts as 
a periodically modulated amplitude of the pitch at $\bar{\omega}$.
This periodically modulated intensity of a pitch is called beat,
indicating some difference between the two pitches exists.
The detection of such beating for a certain pitch 
has commonly been used in tuning the string tensions in piano, guitar, 
to name a few typical instruments.

\section{Classical guitar tuning}
The classical guitar is a plucked string instrument, usually played with fingers.
As an acoustic instrument, it consists of a body with a long, rigid, fretted neck 
and a flat soundboard with incurved sides and a flat back
to which the strings, normally six in number, are attached (see figure 2.2).  
The vibration of the plucked string is transferred through the bridge and amplified by the sound board and resonant body cavity
with the sound being projected from the sound hole.
It belongs to the family of instruments called chordophones (which produce sound from vibrating strings),
and allows the soloist 
to perform complex melodic and polyphonic music pieces in much the same manner as the piano.
 
\begin{figure}[h] \label{TheGuitar}
\includegraphics[scale=0.66]{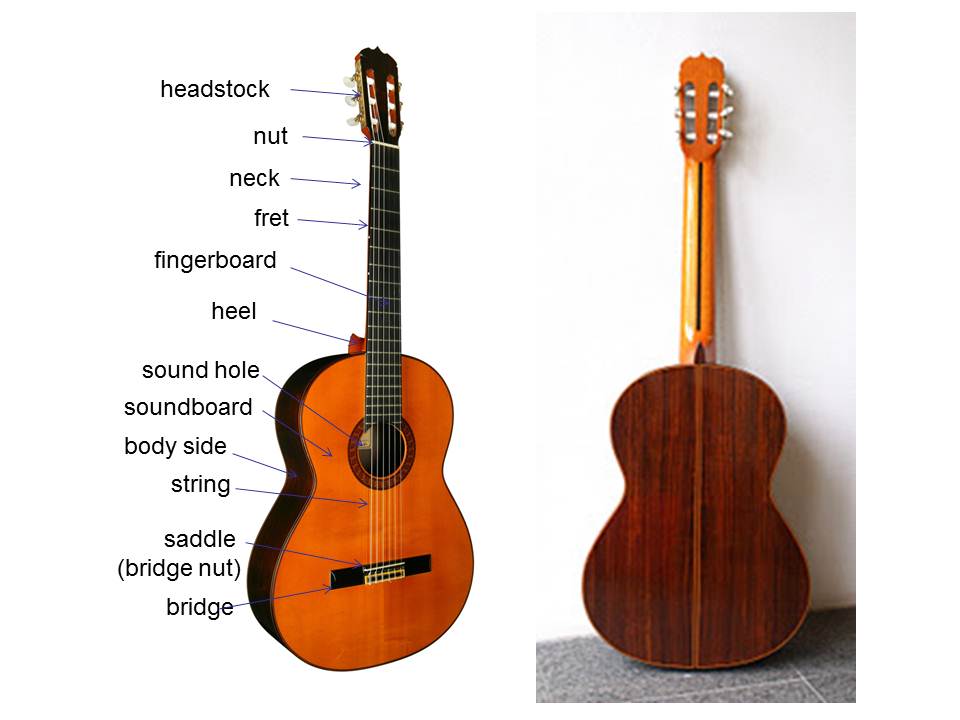}
\caption{Parts of typical classical guitar.  The strings are stretched from their fixed ends at the bridge by 
the tuning pegs on the headstock, pressed against the nut and saddle that determine the scale length.}
\end{figure}

The modern full size classical guitar has a scale length (defined as the length from the nut to saddle) of 
around 650 mm,\footnote{This scale length has remained quite consistent since it was chosen by 
Antonio de Torres (1817-1892)--the
originator of modern classical guitar}
with an overall instrument length of around one meter. \index{Scale length of guitar}
The frets are the raised metal strips embedded along the finger board and placed at points that 
divide the length of string mathematically, such that 
the string's vibrating length becomes that from the fret and saddle when the string is pressed down behind a fret.
Each fret is responsible for producing a different pitch (frequency) and each pitch is spaced 
a half-step (semitone) apart on the 12 tone chromatic scale.  
The ratio of the lengths to saddle (i.e., the $L$ in (1.2)) of two consecutive frets is 
the twelfth root of two ($\sqrt[12]{2} \approx 1.059463$)
according to the equal temperament.  
The twelfth fret is placed at the exact half of the scale length, dividing the string in two exact halves.
Thus, when pressed at the twelfth fret a string will produce a note one octave higher from that of the open string 
(without being pressed anywhere between the nut and saddle so its vibrating length equals the scale length).

A properly constructed guitar should guarantee that each of the six strings to vibrate as a stretched string with two fixed ends,
one at the saddle and the other at the fret where the string is pressed (usually by a left-hand finger), for every fret.
Therefore the spacing between the stretched string and the fretted finger board is gradually increasing 
from the nut toward the bridge, to avoid the buzzing sound arising from the interference from a close-by neighboring fret
when the string is vibrating.

Frets placed at fixed points allow the player to easily and consistently control the string's vibrating length,
without the need of fine adjustment of the exact point (between two frets) 
on fingerboard where the finger is pressing the string.
This enables the player to produce properly sounding chords that require complicated fingering 
for pressing multiple strings simultaneously in polyphonic music playing.
At the same time vibrating strings pressed against frets of hard material induce less damping of the vibrations than 
fingers (of relatively soft flesh) directly.\footnote{However, the equal temperament 
does not yield perfect tuning for perfectly sounding tones that please our ears most; rather it
offers reasonable compromises among the imperfections in all those perfect tuning methods for obtaining equally 
spaced semitones.  For bowed instruments such as violin, which do not have the damping problem due to continued
string stimulation, unfretted fingerboards are used to allow the player more control over fine, subtle changes
in pitch than fretted fingerboards.  Therefore, the music played on violin can be more appealing to our ears
with tones of perfect tuning at any keys. But it is general difficult to play chords having more than a couple of notes
on violin with unfretted fingerboard, because of the requirement of simultaneously adjusting positions of several fingers
with considerable accuracy. }
That is why a (guitar) player with good technique does not pressed the string on the fret but rather behind the fret,
leaving the part of vibrating string between the fret and saddle free of contact, for prolonged ringing of the note.

A common way to tune a guitar is to compare the frequencies of various combinations of 
pairs of the guitar strings that are supposed to produce the same tone when pressed at appropriate frets.
In the standard guitar tuning, the six open strings are tuned at pitches of E2-A2-D3-G3-B3-E4 
from the lowest pitch (low E or E2) to highest (high E or E4). \index{Guitar tuning}
Thus, the B3 string pressed at the 5th fret should produce the E4 tone the same as that of the open high E string.
Similarly, the G3 string pressed at the 4th fret should have the same pitch as 
that of the open B3 string,
and so is the D3 string at 5th fret compared with the open G3 string, the A2 string at 5th fret with the open D3 string,
the low E string at 5th fret with the open A2 string.
When tuning the strings by comparing their difference in pitch, it is not difficult to tell 
whether two strings are different in frequency by as little as a fraction of one hertz,
because the beats can be heard within a second of time interval.
The beats should disappear when the two string frequencies are identical.

Because of its fretted fingerboard, the same musical note can be played on different strings
at different fret positions on a guitar,
unlike the keyboard instrument where a note can only be played on one unique key.
This offers a guitar player the choice of playing a note at different fret on a different string for achieving different
tone color.
At the same time, this give rise to challenges to beginners in finding the correct fret for a given note 
or getting familiarized with a variety of possible fingerings for a given chord, which stays the same on a piano.
Skilled guitar players often produce bright and clear tones of a note in lower positions, and 
softer and richer tones in upper positions of the fingerboard.
Such an enabling property has led guitar players spend lots of time in cultivating their techniques 
for the rewards garnered from rich tone colors.

\newpage
\thispagestyle{empty}

\chapter{Sound Perception}
Perception of complex sound is a process in everyday life and contributes in the way one perceives reality.
There may not be straightforward explanations to sound perception and how it affects 
human beings.
Physics of simple sound can be described as a function of frequency, amplitude, and phase.
Complex sounds can be broken down into a superposition of simple sounds according to Fourier analysis.
Psychology of real-life sound, also termed psychoacoustics, has its own meanings for
pitch, intensity, and timbre.
An interconnection seems to exist between physics and psychology of hearing in terms of 
Fourier decomposed harmonics and their relationship in the frequency domain.
Perception of sound and music is a process for human to distinguish the harmonics in a complex sound wave
under certain circumstances.

\section{Psychoacoustic}
The study of sound perception is called psychoacoustics.
Hearing is not a purely mechanical phenomenon of wave propagation
and air pressure vibration,
but is also a sensory and perceptual event.\cite{moore03}

\subsection{Range of perception}
Human ear can normal hear sounds ranging from 20 Hz to 20 kHz.
(Sounds below 20 Hz are classified as subsonic, 
and those over 20 kHz ultrasonic.)
This upper limit tends to decrease with age, e.g., most adults 
cannot hear sounds above 16 kHz. 
The average frequency of the human voice falls in the range from 120 Hz to
1.1 kHz.
Although the ear itself does not respond to sounds below 20 Hz, 
our ``sense of touch'' can perceive those low-frequency vibrations.
Such a large range of perceivable frequencies might be 
one of the reasons for the fact that
the semi-tone scale in the Western musical notation 
is a logarithmic frequency scale instead of a linear one.
The most sensitive range for the human ear seems to be between 
$2$ and $3$ kHz, which covers the sounds of human scream 
and that produced by opera singers who want to drive the emotion.

Within the octave interval of 1 to 2 kHz, 
the frequency resolution of the ear is about 3.6 Hz,
i.e., changes in pitches greater than 3.6 Hz can be perceived in 
the clinical setting.
However, much smaller pitch differences can be perceived through 
other means such as beating.

Our ear drums can detect sound pressure variations 
as small as $2 \times 10^{-5}$ Pa and as great as or greater than 
$10^5$ Pa (which is about 
1 atmospheric pressure).
Hence sound pressure level is measured logarithmically
with a reference level of $2 \times 10^{-5}$ Pa, 
in units of decibels (dB) 
such that the lower audible limit corresponds to $0$ dB
(cf. table 3.1 for the values of decibel levels for typical sounds). \index{Decibels}
But the upper limit is not as clearly defined because of the question 
of whether it should be the limit at which the ear can be physically
harmed or that may potentially cause the ``noise-induced hearing loss''. 
The sensitivity of our ears also varies tremendously over the audible range.
For example, a 50 Hz sound of 43 dB intensity is perceived 
to be of the same intensity as a 4 kHz sound at 2 dB.

\begin{table}
\caption{Decibel levels for typical sounds}
\begin{center}
\begin{tabular*}{0.75\textwidth}{@{\extracolsep{\fill}} lcc}
\hline
\hline
\\ 
  Source of sound & Decibel level (dB)  & Intensity ($W/m^2$)  \\
\hline
\\
  Threshold of hearing & $0$ & $10^{-12}$  \\
  Breathing & $20$ & $10^{-10}$ \\
  Whispering & $40$ & $10^{-8}$  \\
  Talking softly & $60$ & $10^{-6}$  \\
  Loud conversation & $80$ & $10^{-4}$  \\
  Yelling & $100$ & $10^{-2}$  \\
  Loud concert & $120$ & $1$  \\
  Jet takeoff & $140$ & $100$  \\
\hline
\hline
\end{tabular*}
\end{center}
\end{table}

\subsection{Frequency response over audible range}
Our sensitivity to sounds of different frequencies varies tremendously over the audible range.
Illustrated in table 3.2 is the sound intensity level required for perceiving sounds at various frequencies 
of equal loudness.

\begin{table}
\caption{Sound intensity level required to perceive sounds
at different frequencies to equal loudness}
\begin{center}
\begin{tabular*}{0.75\textwidth}{@{\extracolsep{\fill}} lcc}
\hline
\hline
\\ 
  Frequency (Hz) & Decibel level (dB)  & Intensity ($W/m^2$)  \\
\hline
\\
  50 & $43$ & $2 \times 10^{-8}$  \\
  100 & $30$ & $1.0 \times 10^{-9}$ \\
  200 & $19$ & $7.9 \times 10^{-11}$  \\
  500 & $11$ & $1.3 \times 10^{-11}$  \\
  1000 & $10$ & $1.0 \times 10^{-11}$  \\
  2000 & $8$ & $6.3 \times 10^{-12}$  \\
  3000 & $3$ & $2.0 \times 10^{-12}$  \\
  4000 & $2$ & $1.6 \times 10^{-12}$  \\
  5000 & $7$ & $5.0 \times 10^{-12}$  \\
  6000 & $8$ & $6.3 \times 10^{-12}$  \\
  7000 & $11$ & $1.3 \times 10^{-11}$  \\
  8000 & $20$ & $1.0 \times 10^{-10}$  \\
  9000 & $22$ & $1.6 \times 10^{-10}$  \\
  14000 & $31$ & $1.3 \times 10^{-9}$  \\
\hline
\hline
\end{tabular*}
\end{center}
\end{table}

Figure 3.1 shows equal-loudness contours used to indicate the sound pressure (dB),
over the range of audible frequencies, as perceived being of equal loudness.
Equal-loudness contours were first measured using headphones (by Fletcher and Munson in 1933).
As a psychological quantity, loudness is difficult to measure and requires to take average over many tests.
The equal-loudness curves derived using headphones are valid only for the special case of what is called `side-presentation', 
which is not how we normally hear.
Real-life sounds arrive as planar waves, if from a reasonably distant source. 
If the source of sound is directly in front of the listener, 
both ears receive equal intensity.
But at frequencies above $1$ kHz, the sound that enters the ear canal is partially reduced by the
``masking effect of the head'', and is also highly dependent on reflection off the pinna (outer ear).
Off-center sounds result in increased head masking at one ear, and subtle changes in the effect of the pinna, especially at the other ear.
This combined effect of head-masking and pinna reflection is quantified in a set of curves in 
a three-dimensional space referred to as ``head-related transfer functions''.
Frontal presentation is now regarded as  preferable when deriving equal-loudness contours.  

\begin{figure}[h] \label{EqualLoudness}
\includegraphics[scale=0.56]{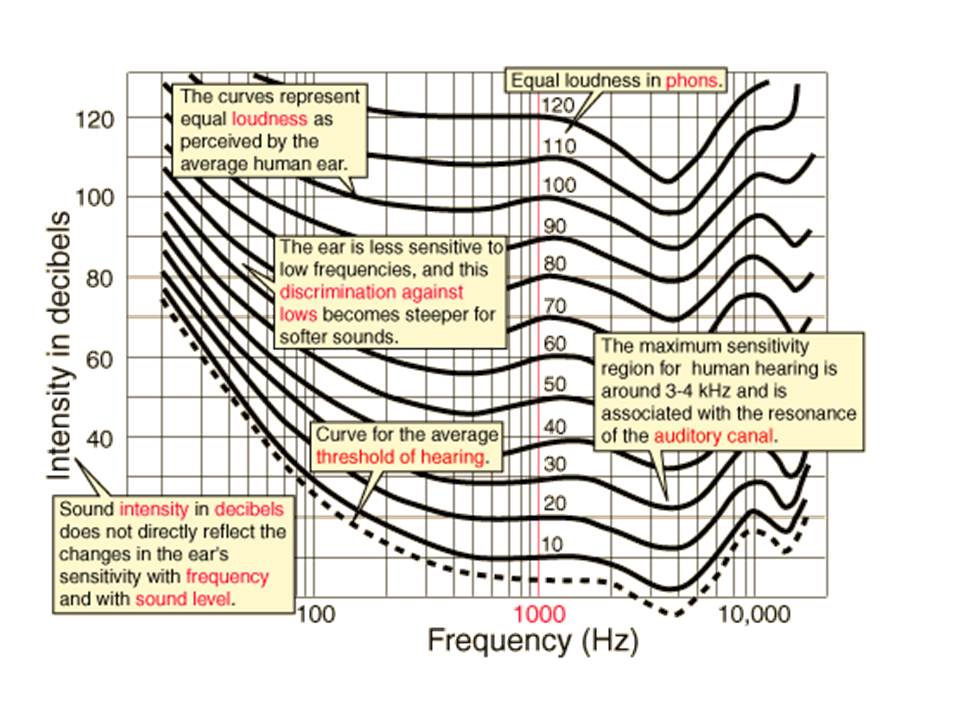}
\caption{Equal-loudness curves.}
\end{figure}

It appears that the human ear is most sensitive between $2$ kHz and $5$ kHz, largely due to the resonance of 
the ear canal and the transfer function of the ossicles of the middle ear.

\subsection{Critical bands}
When sound enters the ear, it ultimately causes vibrations on the basilar membrane within the inner ear.
Different frequencies of sound cause different regions of the basilar membrane and its fine hairs to vibrate.
The region difference on the basilar membrane is how the brain discriminates between various frequencies.
When two frequencies are close together, there is an overlap of response regions on the basilar membrane.
If the frequencies are nearly the same, they cannot be distinguished by the brain as separate frequencies.
Instead an average (fused) frequency (such as $\bar{\omega}$ in (2.2)) is heard, as well as the beats (of a frequency 
$\Delta \omega$ in (2.2)).

\begin{figure}[h] \label{CriticalBand}
\includegraphics[scale=0.64]{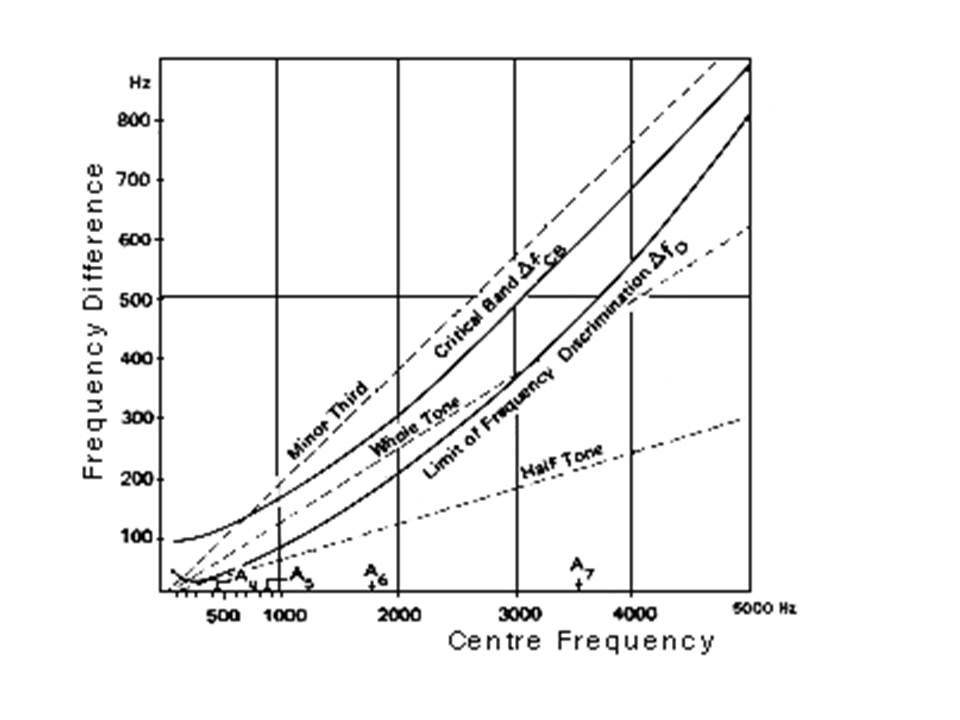}
\caption{Critical bandwidth $\Delta f_{CB}$ and limit of frequency discrimination $\Delta f_D$ 
 as a function of the central frequency of a two-tone stimulus.}
\end{figure}

However, if the difference between the two frequencies increases,
there will come a point at which the two frequencies are still indistinguishable, but the beat frequency will be 
too high to make out such that only a roughness to the sound---dissonance---can be perceived,
indicating that both frequencies are activating the same part of the basilar membrane.
This dissonance will continue with increasing the frequency difference until finally the two frequencies become
distinguishable as separate frequencies with a difference $\Delta f_D$.
At the point where the ``roughness'' disappears, the frequency separation is equal to the so-called critical bandwidth $\Delta f_{CB}$ 
and the two frequencies activate different sections of the basilar membrane; 
further increasing the frequency difference will  produce less and less dissonance.
The two frequencies are said to be within a critical band on the basilar membrane
when they are close enough to cause the beating and roughness on top of the perceived average frequency.
For much of the audible range, 
the critical bandwidth ($\Delta f_{CB}$) around some ``central frequency'' 
is typically stimulated by frequencies within about $15\%$ 
of that central frequency (as shown in figure 3.2). \index{Critical bandwidth}

The usable tones in musical scales (for producing consonance) are actually limited by the effect of critical bands.
If two tones with frequencies that are too close to each other that stimulate areas within the 
same critical band on the basilar membrane, they will produce either noticeable beats or other dissonance 
(e.g., unpleasantly rough sound) undesirable in music rather than two distinguishable tones.

\subsection{Frequency relationship among the musical notes}
The particular frequency relationship among the notes in musical scale is not random.  
Many notes (or tones) in the musical scale (e.g., those with intervals greater or equal to that of minor third)
sound particularly good when played together at the same time.
The pleasant sounding combinations of frequencies yields consonance, which is the opposite of dissonance.
\index{Consonance} \index{Dissonance}
It has been known that any tone played by a musical instrument contains a series of overtone harmonics with
frequencies equal to the fundamental frequency multiplied by integers,
namely a musical tone can be expressed as a Fourier series similar to that given by (1.2).
Actually what makes two tones sound consonant is that they have many overtone harmonics sharing the same
frequencies.
For example, two tones with one tone of the frequency twice as that of the other always sound good when played together;
in fact it is so good that they sound ``the same'' (because every overtone of the octave tone has the same frequency
as one of those overtones of the other tone.)

In developing the musical scale, it becomes to decide how many different tones to incorporate 
and how far apart in frequency they should be.
There are two reasonable constraints that the those tones should be fairly evenly spaced 
and that they should sound good when played together.
By plucking a string under tension with a movable bridge between the two fixed ends,
the Greek mathematician, Pythagoras, found that it sounded good 
when the ratio of the two lengths from the bridge to an end is 
simply $2:1$, and good consonance were also obtained when the string length ratio becomes 
$3:2$ and $4:3$.  
What Pythagoras found is that consonance is obtained from two tones of frequencies 
with small integer ratios.\footnote{However, as the integers becomes larger the number of shared overtones decreases.
For example, the case of $3:2$ with one tone of 100 Hz and the other 150 Hz has the frequency spectra as

100Hz, 200Hz, 300Hz, 400Hz, 500Hz, 600Hz, ...

150Hz, 300Hz, 450Hz, 600Hz, 750Hz, 900Hz, ...

\noindent
where the second series only have every other overtone harmonic sharing every third overtone harmonic of the first series.}
Thus he constructed the Pythagorean Scale based on $1$, $4/3$ (perfect fourth), $3/2$ (perfect fifth), and $2$ (octave)
with frequency ratios as 1, 9/8, 81/64, 4/3, 3/2, 27/16, 243/128, and 2 for do, re, mi, fa, so, la, ti, and do, 
respectively.\footnote{Following the similar logic of Pythagoras, a (major) pentatonic (five-note) scale can be constructed 
with frequency ratios as 1, 9/8, 4/3, 3/2, 27/16, and 2 (for do, re, mi, so, la, and do), 
where the notes having semitone intervals in the heptatonic (seven-note) scale such as ``fa'' and ``ti'' are left out.
The major pentatonic scale is the basic scale of the music in China as well as in many other
cultures.}.  \index{Pentatonic scale}

Ptolemy, a Greek astronomer, later found an additional consonance in the frequency ratio of $5:4$.  
He noticed that groups of three frequencies sounded particularly good together when their ratios to each other were
$4:5:6$ (which is actually the major triad ``do-mi-so'').  
So he constructed the so-called Just Intonation Scale with the corresponding intervals listed in table 3.3.  

\begin{table}
\caption{Just Scale interval ratios and common names}
\begin{center}
\begin{tabular*}{0.75\textwidth}{@{\extracolsep{\fill}} lcc}
\hline
\hline
\\ 
Frequency ratio & Interval & Interval name \\
\hline
\\
 $2/1$ & C $ ->$ C & Octave  \\
 $3/2$ & C $ ->$ G & Perfect fifth \\
 $4/3$ & C $ ->$ F & Perfect fourth  \\
 $5/3$ & C $ ->$ A & Major sixth  \\
 $5/4$ & C $ ->$ E & Major third  \\
 $8/5$ & E $ ->$ C & Minor sixth  \\
 $6/5$ & A $ ->$ C & Minor third  \\
 $9/8$ & C $ ->$ D & Major second  \\
 $16/15$ & B $ ->$ C & Minor second  \\
\hline
\hline
\end{tabular*}
\end{center}
\end{table}

But those scales based on frequencies with perfect small integer ratios cannot have uniformly distributed frequency spacing,
which is needed for transposing songs to different keys.  
Thus the 12-tone equal temperament scale became a universally accepted compromise between
the consonance of perfect intervals and the equal intervals, 
with the spacing between its notes just to be very close to that of the consonant intervals.
Other equal interval temperaments with different numbers of tones from twelve has not been found to work out as nicely.

On the other hand, for tones in a musical scale to produce consonance when played together, 
their frequencies should neither be close enough to 
cause beats nor be within the critical band.  
Many of the overtones of these two tones should coincide
and most of the ones that do not should neither cause beats nor be within the critical band.
Based on figure 3.2 the minor third (or major sixth) is the smallest interval between tones for consonance.

\section{Restoration of the missing fundamental}  \index{Missing fundamental restoration}
Because the vibrating objects usually emit sounds that contain multiple 
harmonic components (see, e.g., figure 1.2),
our brains respond to such harmonic sounds with synchronous neural firings,
creating a neural basis for the coherence of these sounds.
In fact, our brains are so attuned to the overtone series that if we encounter 
a sound that has all of the components except the fundamental,
we would still perceive the sound as the original note with the fundamental pitch.
For example, a sound composed of harmonic overtones at 200 Hz, 300 Hz, 400 Hz, and 500 Hz 
is still perceived by our brains as having a pitch of 100 Hz,
despite that the fundamental mode of 100 Hz is left off.  
This phenomenon is called ``restoration of the missing fundamental.''\cite{levitin07, roederer08}
It has a practical application in our telephone communication system,
which does not have the bandwidth large enough to cover the entire vocal range of 
male and female voices.  
Human voices range from 80 Hz to 1100 Hz (male: 80 - 440 Hz and female: 230 - 1100 Hz),
whereas most telephones transmit from about 300 Hz to 3400 Hz.
Yet when we talk on the phone, we still perceive the other's voice quite clearly 
and distinctively for the most part, although what we are actually hearing are just 
the overtones.  

Mathematically, a pressure signal $f(t)$ expressed as a summation of harmonic components
(in terms of a  Fourier series)
\begin{eqnarray} \label{harmonic_sum_eq}
f(t) = A_1 \cos(2 \pi 100 t) + A_2 \cos(2 \pi 200 t) + A_3 \cos(2 \pi 300 t)  + \\ 
\quad \quad \quad \quad + A_4 \cos(2 \pi 400 t) + A_5 \cos(2 \pi 500 t) \nonumber  \, 
\end{eqnarray}
exhibits the (common) periodicity of a frequency at 100 Hz even when $A_1 = 0$.
In other words, $f(t) = 0$ always occurs at $t = 0.005$ and $0.001$ but not 
at $t = 0.0025$ corresponding to a frequency of 200 Hz or $t = 0.00165$ 
for a frequency of 300 Hz.
Such a common periodicity reflects the fundamental frequency inherent in all the 
harmonic components, and can be picked up by our brain even when the fundamental 
component at frequency of 100 Hz is missing (i.e., $A_1 = 0$).
Thus, in processing a sound signal, our brains must have been 
identifying the periodicity of it for the pitch.
The expression of Fourier series is convenient for mathematicians to do their analysis 
but is not the way our brains sense the pitch of sound.
In fact, our auditory system is capable of identifying 
the repetition rate---periodicity---of a complex sound.
As a consequence, a missing fundamental tone with a respective frequency,
called virtual pitch, can be perceived in the brain through the ``subjective pitch detection'',
\index{Subjective pitch detection} 
while not being directly perceived by the ear.

\section{Feature extraction and integration} \index{Feature extraction and integration}
Generally speaking, our perception is a process of inference, and involves 
an analysis of probabilities.
The brain's task is to determine what the most likely arrangement of objects in the 
physical world is, given the particular pattern of information that 
reaches the sensory receptors---the retina for vision, the eardrum for hearing.
More often than not, the information we receive at our sensory receptors is
incomplete or ambiguous.
The voices we listen to are typically mixed with other voices such as wind, footsteps, etc.
Our brains can extract basic, low-level (i.e., elemental building-block) 
features from the music, using specialized
neutral networks that decompose the signal into information about pitch, timbre, spatial location,
loudness, reverberant environment, tone durations, and the onset times 
for different notes and for different components of tones. 
The low-level processing of basic elements occurs in the peripheral and phylogenetically older
parts of our brains,
whereas high-level processing---feature integration---occurs 
in more sophisticated parts of our brains 
that take neural projections from the sensory receptors and from a number of low-level
processing units to combine low-level elements into an integrated representation.
It is the high-level processing where it all comes together, 
where our minds come to an understanding of form and contents.

At the same time as feature extraction is taking place in the cochlea auditory cortex,
brain stem, and cerebellum,
the higher-level centers of our brain are receiving a continuous stream of information 
about what has been extracted so far;
this information is continually updated, and typically rewrites the older information.
As our centers for higher thought---mostly in the frontal cortex---receive these updates,
they are trying to predict what will come next in the music, 
based on what has already come before in the piece of music we are hearing
as well as our knowledge about the music.
These frontal-lobe calculations and expectations can influence 
feature extraction of the lower-level modules,
causing us to misperceive things by resetting some of the circuitry in neutral networks 
for perceptual completion\footnote{Perceptual completion is a phenomenon
when our perceptual system completes or ``fills in'' information
that is not actually there.
This somewhat subconscious capability of our perceptual system for restoring missing information
would help us make quick decisions in threatening situations 
for survival purposes, because much of what we see and hear 
almost always contains missing information.} and other illusions.\cite{levitin07}
\index{Perceptual completion}

\section{Music and life}
There have been theories attempting to explain the origin of music.  One derives music
from the inflections of speech, another from mating calls,
a third from the cries of battle and signals of the hunt,
a fourth from the rhythms of collective labor.
Some people attribute the rise of music to the imitation of nature, 
others connect it with the play impulse,
with magic and religious rites, or with the need for emotional expression. 
One common factor is that music relates to the profoundest experiences of the individual and the group.
Underlying all is the fact that man possesses in his vocal cords a means of producing song,
in his body an instrument for rhythm,
and in his mind the capacity to imagine and perceive musical sounds.

Life in human society is filled with emotion.  
Speech and body movement alike are intensely expressive.
Speech heightened by musical inflection becomes song.
Body movement enlivened by musical rhythm becomes dance.  
Song and dance---that is, melody and rhythm---constitute the primal sources of music.

In a sense we may speak of music as a universal language,
one that transcends the barriers men put up against each other.
Its vocabulary has been shaped by thousands of years of human experience; 
its rhetoric mirrors man's existence, his place in nature and in society.
It has been proven that music influences people with effects that are
instant and long lasting.
Music may be considered to link all of the emotional, spiritual,
and physical elements of the universe.

\newpage
\thispagestyle{empty}


\chapter{Basic Elements of Music}
A musical unity comes from an (intelligent) organization of 
a few basic elements such as rhythm, meter, tempo, melody, harmony, timbre, dynamics, etc.
These basic elements form the structure and framework for which music is 
built up.\cite{machlis63}
Understanding these element allows one to study, play, compose, reharmonize, 
and work with other musicians in a common language.

\section{Melody: musical line}
Melody is that element of music which makes the widest and most direct appeal.
It has been called the soul of music.
It is general what we remember and whistle and hum.
We know a good melody when we hear it and we recognize its unique power to move us,
although we might not be able to explain wherein its power lies.

Yet, a melody is nothing more than a linear succession of tones perceived by the mind as a single entity.
In order to perceive a melody as an entity we must find a significant relationship among its constituent tones.
This places a certain responsibility upon us---the listener.  
We derive from the succession of tones an impression of conscious arrangement: 
the sense of a beginning, a middle, and an end.
It is a combination of pitch and rhythm, simply put.
A good melody has something inevitable; it possesses a distinctive profile, a quality of aliveness.

We perceive tones not separately but in relationship to each other within a pattern.
Tones move up and down, one being higher or
lower than another in ``musical space''.  
They also move faster or slower in time, one claiming our attention for a longer or shorter duration than another.
From the interaction of the two dimensions---musical space and time---emerges 
the total unit which is melody.
This is the musical line, or curve, which guides our ear through a composition.
It is the plot, the theme of a musical work, the thread upon which hangs the tale.

The characteristics of melody include range, shape, and movement.

The range of a melody is the distance between its lowest and highest tones. 
Singers refer to an arrangement being in a low, medium, or high range,
meaning that the notes focus on those scale pitches.

The shape of a melody line refers to the literal geometric line that could be made 
if the notes on a music score were joined together as in a dot-to-dot puzzle.
Notes that ascend up the scale take on an upward shape, while 
phrases that descend are shaped in a downward motion.  
It the phrase stays within a narrow range, the shape is wavelike.

The movement of a melody can be either conjunct or disjunct.
When the melody moves stepwise and is connected, its movement is termed conjuncted.
Melody that leaps from pitch to pitch with no natural connection or flow is 
said to be disjunct.  \index{Conjunct melody} \index{Disjunct melody}
Our brains feel comfortable with a melody having a lot of stepwise motion
consisting of adjacent tones in the scale, i.e., conjuncted melody.
If the melody makes a big leap---as in a disjunct melody---there is 
a tendency for it to return to the 
jumping-off point.  Because our brains would anticipate 
that the leap is only temporary, and tones that follow need to bring us closer and 
closer to the starting point, or harmonic ``home''.
Thus, after the melody makes a large leap, either up or down,
the next note usually changes the direction of melodic motion.

Melody is structured much like sentences in a spoken language.  
For instance, a phrase in music is a unit in its entirety within the larger structure of the song.
There are also the cadence and the climax.  
A cadence is a melodic or harmonic configuration that creates a sense of repose or resolution  
such as a final ending to a musical section, 
and a climax is the most intense and emotional part of a phrase
(though not necessarily the highest or the loudest tone). \index{Cadence} \index{Climax}

It is the discrete set of tones called the musical scale that forms the building blocks of melody. \index{Musical scale}
As explained in sections 2.1 and 3.1.4, these tones in a musical scale have well-defined relationship 
in terms of their frequency ratios (such as those in table 2.1) which closely associate with 
the way our ears perceive the sound.  
Here we attempt to discuss the relationship among these seven tones from the perspective of their importance 
in music making with the diatonic scale (i.e., do-re-mi-fa-so-la-ti-do).\footnote{In tonal music
of the 18th and 19th centuries, the tonic center was the most important of all the different tone centers
which a composer often used in a piece of music.
Most tonal music pieces begin and end on the tonic, usually with modulation to the dominant (the fifth note above or
the fourth note down with respect to the tonic note) in between.}
Tonality is a hierarchy of tones (pitches) where one pitch (i.e., the tonic) has prominence over the others, i.e., some are more stable or final sounding than others, 
invoking a feeling of varying amounts of tension and resolution.
In the major scale, the most stable tone is the first degree, called the tonic, 
toward which other tones tend to move with the seventh degree---leading tone---having the
strongest such motion tendency while the fifth degree--dominant the weakest.
That is why we do not feel uneasy---unresolved---if a song ends on the fifth scale degree
instead of on the tonic.
Thus, in a hierarchy of importance among scale tones, tonic is the key note, 
around which a fifth degree above is the dominant and a fifth degree below the subdominant.
The mediant is midway between tonic and dominant (or the middle note of the tonic triad); 
the submediant is midway between tonic and subdominant (or the middle note of the subdominant triad).
The supertonic and leading tone are a step away on either side of the tonic.
Table 4.1 lists the names of scale degrees of a diatonic scale according to their importance.
Interestingly,
the supertonic also comes from a fifth above the dominant; then a fifth above the supertonic is 
the submediant, and a fifth above the submediant is the mediant with 
the leading tone situated yet a fifth above the mediant.

\begin{table}
\caption{The names of the scale degrees of a diatonic scale.}
\begin{center}
\begin{tabular*}{0.75\textwidth}{@{\extracolsep{\fill}} lcc}
\hline
\hline
\\ 
  Tone & Scale degree  & Scale degree name  \\
\hline
\\
  do & first degree (unison)  & Tonic  \\
  so & fifth degree  & Dominant \\
  fa & fourth degree & Subdominant  \\
  re & second degree & Supertonic  \\
  la & sixth degree & Submediant  \\
  mi & third degree & Mediant  \\
  ti & seventh degree & Leading tone  \\
\hline
\hline
\end{tabular*}
\end{center}
\end{table}

Among the seven tones, the tonic, dominant, and subdominant are called the principal scale tones because they are 
in the most influential positions in music making. \index{Principal scale tones}
The principal scale tones serve as the roots for principal chords. \index{Principal chords}
The supertonic, submediant, and mediant belong to the subordinate scale tones, 
corresponding to the roots for subordinate chords.
The leading tone has a special effect that tends to move toward (or resolve to) the tonic.

The leading tone, submediant, subdominant, and supertonic tones are also called active tones, 
because they have the tendency to resolve in 
a special direction toward (or be attracted to)
the nearby stable (or inactive) tones (such as those tones in the tonic triad) in a melody. \index{Active tone} \index{Stable tone}
For example, the leading tone tends to move up toward the tonic, while the submediant toward the dominant,
the subdominant toward the mediant.
The supertonic tone can either move down toward the tonic or up toward the mediant, for its ``distance'' to
both are ``equal''.
Therefore, the active tones bring motion and liveliness in the melody line, while the inactive tones usually conclude 
the musical phases leaving a settled, rest feeling to the listener.

\section{Harmony: musical space}
To the movement of the melody, 
harmony adds another dimension---depth.
``Harmony is to music what perspective is to painting.''
It introduces the impression of musical space,
and clarifies direction and creates meaning.

Harmony pertains to the movement and relationship of chords.
A chord may be defined as a combination of tones occurring simultaneously and conceived as an entity.
Just as the vaulting arch rests upon columns, 
the melody unfolds above the supporting chords, the harmony.
Melody constitutes the horizontal aspects of music; harmony, the vertical.

\subsection{The function of harmony}
The chords may serve as the framework of a composition, forming the substructure that holds it together.
They have meaning only in relation to other chords; namely, only as each leads to the next.
Harmony therefore implies movement, progression.
In a sense, harmony denotes the over-all organization of tones in a musical work in such a way as to 
achieve order and unity.

The most common chord is a certain combination of three tones known as a triad.  \index{Triad}
Such a chord may be built by combining the first, third, and fifth degrees of the do-re-mi-fa-so-la-ti-do scale:
do-mi-so.
A triad may also be built on the second degree (re-fa-la); on the third degree (mi-so-ti); and similarly 
on each of the other degrees of the scale (as a chord is build with the interval of thirds). 
It always starts from one tone (or root) and ends at the fifth tone above.
If the interval between the root tone and the middle tone is a major (or minor) third, it is called a major (or minor) chord.

Although the triad is a vertical block of sound, its three tones often appear horizontally as a melody.
 The melody and harmony actually do not function independently of one another;
on the contrary, the melody implies the harmony that goes with it, and each constantly influence the other.

\subsection{Active and rest chords}
Music is an art of movement.  A purposeful movement means there must be a goal.  
The progressions of chords follow certain logic that art imposes upon nature. 

Purposeful movement implies a central point of departure and return.
A point in the final tone of melodies can typically be noticed.  This is the ``do'' (as the tonic)
which comes both first and last in the do-re-mi-fa-so-la-ti-do scale.  
The triad on ``do'' (do-mi-so) is the ``I'' chord or Tonic, which serves as the chord of rest.
But rest has meaning only in relation to activity.  
The active chords seek to be resolved in the rest chord. \index{Active chord} \index{Rest chord}
This striving for resolution is the dynamic force in the music.
It shapes the forward movement, imparting direction and goal.

The fifth step of the do-re-mi-fa-so-la-ti-do scale is the chief representative of the active principle.
The active triad rooted on ``so'' (so-ti-re) is the ``V'' chord or Dominant,
which seeks to be resolved in the restful I chord rooted on ``do''.
Dominant moving to Tonic constitutes a compact formula of activity completed,
of movement come to rest.
Therefore the Dominant-Tonic progression shapes the most common final cadence in music. 

The triad built on the fourth scale step ``fa'' (fa-la-do) is called the ``IV'' chord or Subdominant. 
This too is an active chord, but less so than the Dominant.
The progression IV-I creates a less decisive cadence than the V-I progression.

These three triads, the Tonic, Dominant, and Subdominant, usually suffice to harmonize many melodies.

\subsection{Consonance and dissonance}
Harmonic movement in music is generated by the tendency of the active chord to be resolved in the chord of rest.
This movement receives its maximum impetus from the dissonance, which creates a feeling of restlessness and activity,
whereas consonance is relaxation and fulfillment.
The dissonant chord creates tension, and the consonant chord resolves it. \index{Consonance} \index{Dissonance}

A dissonant chord can be obtained by stacking another third tone on top of the existing triad, i.e., adding a seventh note 
from the root.  This produces a four-note chord called a ``seventh chord''.
Because there is an interval of major or minor second due to the tone of seventh degree, 
the seventh chord sounds dissonant.
This seventh tone destabilizes the triad, allowing the composer to emphasize movement in a given direction.
Such a dissonant chord is often used in the V chord as the Dominant seventh.

A single note by itself cannot be dissonant, but it can sound dissonant against
the background of certain chords,
particularly when the chord implies a key that this single note is not part of.
Two notes can sound dissonant together, when both played simultaneously
or in sequence, if the sequence does not conform to certain musical idioms.
Chords can also sound dissonant, especially when they are drawn from outside
the key that has been established in the musical sequence.

\section{Rhythm and meter: musical time}
Rhythm denotes the controlled movement of music in time.
It is the element of music most closely allied to body movement, to physical action, and is called 
the heartbeat of music, the pulse that betokens life.
It reflects the regularity with which we walk, breathe, as well as the periodicity in numerous cyclical natural phenomena.
Actually making music consists of the coordinated, rhythmic movements of our body 
that are subsequently transmitted to a musical instrument.
At the neural level, playing an instrument requires the orchestration of regions in our 
primitive, reptilian brain---the cerebellum and the brain stem---as well as higher cognitive systems such as the motor cortex (in the parietal lobe)
and the planning regions of our frontal lobes in the most advanced region of 
the brain.\cite{levitin07}

Rhythm releases our motor reflexes,
causing people to fall in step when the music is on, to nod or tap with the beat.
It springs from the need for order inherent in the human mind.
Upon the tick-tock of the clock we automatically impose a pattern,
and we organize our perception of time by means of rhythm.

Rhythm is marked by the regulated succession of opposite elements,
the dynamics of the strong beat and weak beat, the played beat and the inaudible but implied rest beat,
the long note and short note.
As well as perceiving rhythm we must be able to anticipate it.
This depends upon repetition of a pattern that is short enough to memorize. 

Meter pertains to the organization of musical time, constituting the basic scheme of accents in a composition;
specifically, to the arrangement of musical beats in units of time known as ``measures''.
Each measure contains a fixed number of beats and are marked off by the recurrence of a strong accent (stressed beat).
In other words, recurring beats are organized into regular groups (measures) by accenting the first beat of each group.
These metric groupings are divided into measures by barlines. \index{Measures} \index{Barlines}
For example, duple meter is the alternation of strong and weak beats, where there are two beats per measure.
Triple meter has three beats per measure, with the first beat usually being the strong beat.
Quadruple meter has four beats per measure.  The first and third beats are stressed,
with first beat usually being stronger than the third.

The standard way for the time signature to appear in music is after the key signature with 
one number directly above the other, e.g., 2 over 4 as if it were a fraction. \index{Time signature}
The upper number tells how many notes of the type represented by the lower number 
there are in each measure.
For instance, in 6 over 8 time the lower number 8 would be the denominator of the fraction 
one eighth, so there are 6 eighth notes in a measure of 6 over 8.
Likewise, in 4 over 4 time there are 4 quarter notes in a measure, 
2 over 16 time has 2 sixteenth notes in a measure, etc.
The next thing to determine about a meter is what kind of note represents one beat,
and how many beats are in a measure.  This brings up the concept of simple time and compound time.

Simple and compound time refers to the division of the beat. 
In simple time the beat is divided into 2 equal parts, whereas in compound time 
the beat is divided into 3 equal parts.
Meters where the upper number is both larger than and divisible by 3 (6, 9, 12, etc.) are compound,
whereas those with the upper number being an even number not divisible by 3 (2, 4, 8, etc.) are simple.
For example, the upper number of 2 over 4 (duple meter) is not divisible by 3, so the meter is simple.
The upper number of 6 over 8 is larger than and divisible by 3, so it is compound time
(where the beat is divided into 3 equal parts in this compound time, so there are 3 eighth notes in each beat.)
The upper number of 3 over 4 (triple meter) is divisible by 3, but is not larger than 3, so the meter is simple
(where the upper number is the number of beats in a measure and the lower number is the value of the note 
that represents the beat.)

For the convenience of identifying the time (or meter) signature,
it is customary to beam any notes faster than the quarter note together as nearly as possible by beat.
For example in 6 over 8 time the eighth notes will be beamed together in 2 groups of 3 each,
while in 3 over 4 time the eighth notes will be beamed in 3 groups of 2 each.

Noteworthy here is that the even number times (such as $4/4$ and $2/4$) 
are easy to walk to, dance to, or march to 
because you always end up with the same foot hitting the floor on a strong beat.
Three-quarter is less natural to walk to;
you never see a military marching that follows a $3/4$ time.

\section{Tempo: musical pace}
Meter tells us how many beats are in the measure, but it does not tell us whether these beats occur slowly or rapidly.
The tempo, which determines the speed of the beats, provides the answer to this matter.
In music tempo is an important factor in conveying emotion,
namely fast tempos tend to be perceived as happy and slow tempos as sad.
On a music sheet, the tempo is indicated by the composer typically in terms of 
the number of quarter notes (or beats) per minute.
It may also be expressed verbally with such words as ``Largo'' (slowly), ``Allegro'' (fast), etc.
\index{Largo} \index{Allegro}
However, musical tempo is a much more elastic concept than physical time.
The movement of a musical piece is a live thing, now pressing forward, now holding back.
No artist playing the same work on two occasions could possibly duplicate the tempo throughout,
with its infinite nuances, its subtle accelerations and retardations.\footnote{In fact,
most people cannot detect a variation of tempo by less than 4 percent,
although professional drummers may be able to detect smaller tempo variations
due to their job requirements for more sensitive to tempo than other musicians.\cite{levitin07}}
The musician adheres to the beat in order to project a clear picture of the meter, rhythm, and tempo; 
he also departs from the beat in order to achieve the necessary suppleness of movement.
In this way he breathes life into what would otherwise be a mechanical thing,
transforming the music into a free flow of thought and feeling.

\section{Timbre: musical color}
As discussed in section 1.3 a tone produced by different instruments will sound differently 
due to the waveform differences.
Such difference lies in the timbre (or color) of each instrument, which imparts to 
the tonal image its special and inalienable character.\footnote{Careful observation 
shows that the waveform
of sound from a musical instrument also changes after its initiation, with a ``noisy'' 
beginning right after the pluck, hit, strum, bowing, or blowing followed by a more stable 
orderly pattern of overtones from the instrument resonance.  
Although many sound components in the initial noisy (attack) phase decay quickly, they play 
an indispensable role in defining the timbre of the instrument.\cite{levitin07}}
For a composer to write idiomatically for an instrument,
it means to function musically within its limitations---perhaps even to transmute 
these into fresh sources of beauty.
Each instrument has the limit of its range---the distance from its lowest to its highest tone.
There are also the limits of dynamics---the degree of softness or loudness beyond which it cannot be played.
Yet tone color of an instrument is not something that is grafted on to the musical conception;
it is part and parcel of the idea, as inseparable from is as are its harmony and rhythm.
Timbre is more than an element of sensuous charm that is added to a work; 
it is one of the shaping forces in music.

In scientific terms, however, the meaning of ``timbre'' takes an unusual definition as 
that it ``is everything about a sound that is not loudness or pitch'' 
(according to the Acoustical Society of America).  
This indicates the complexity of the subject of music; 
even many musicologists and scientists could not agree about the definite meanings
of some of the musical terms.

\section{Dynamics: degree of loudness or softness in music}
The loudness is related to the amplitude of a vibration or the sound pressure (or intensity) level (in terms of decibels). 
In music the effect of loudness or softness is more or less in a relative sense; 
dynamics do not indicate the specific decibel levels.
A gradual increase in loudness is called ``Crescendo'', usually creating excitement, particularly when the pitch rises, too.\index{Crescendo}
A gradual decrease in loudness is called ``Decrescendo'', usually conveying a sense of calm.
\index{Decrescendo}
When notating music, composers have traditionally used Italian words
such as ``forte'' (or ``$f$'')  for loud and ``piano'' (or ``$p$'') for soft, to indicate dynamics.
\index{Forte ``$f$''} \index{Piano ``$p$''} 
Sudden changes in dynamics are notated
by an ``s'' prefixing the new dynamic notation (e.g., ``sf'', ``sff'',  ``sp'', etc.),
where the prefix is called ``subito''---an Italian word which translates into ``suddenly''.
\index{Subito}
A forceful, sudden accent is indicated by ``sforzando'' or ``sforzato'' 
with an abbreviation as ``sfz''.  \index{Sforzando (or sforzato)}
Regular forzando ``fz'' indicates a forceful note, but with a slightly less sudden accent.
\index{Forzando}

Another aspect of dynamics is associated with the execution of events in a given music piece,
either stylistic (``staccato'', ``legato'', etc.)  
or functional (relative velocity).   
Staccato, marked as musical articulation, is often referred to as ``separated'' or ``detached'' 
rather than having a defined, or numbered amount
by which the separation or detachment is to take place. \index{Staccato}
Legato, on the other hand, means ``smooth'', ``connected''. \index{Legato}

\newpage
\thispagestyle{empty}

\chapter{Theory of Harmony}
A harmony is an abstract sense that can be identified by 
a set of tones for forming a musical scale or chord.
Two tones define an interval.  
The most prominent interval is the octave, corresponding to a frequency ratio of 2:1.
Because the two tones that are an octave apart sound the ``same'', 
they are assumed to be octave equivalent and assigned the same names.
With the division of an octave into 12 semitones, intervals may also be defined by 
number of semitones between the two tones.

A chord is a group of simultaneously sounding tones,
while a scale is a set of musical tones usually sounding consecutively.
Both can be identified by the number of semitones in the harmony.

In music, harmony is the sound of simultaneous pitches (tones, notes).
It involves chords---conceived combination of simultaneous tones as an entity, chord construction, 
chord progression, and principles that govern chord connections.\cite{schoenberg83}

\section{Chord construction} \index{Chord construction}
The most commonly used chords are that consist of three tones known as triads based on 
a seven-tone scale. \index{Triad}
Classical triads are built from major and minor thirds, i.e., 
the interval between successive pairs of tones contains 3 or 4 semitones.
In other words, chords are constructed in the ascending direction with every other scale tone 
above the given root.\footnote{The use of thirds in a triad ensures consonant sounding chord 
because none of the intervals will have two tones with their frequency difference less than the critical bandwidth}
In its ``root position'',
the major triad has the $\{0, 4, 7\}$ semitones, and the minor triad has the $\{0, 3, 7\}$ semitones. 
Thus both major triad and minor triad contain a (perfect) fifth tone, yet they differ by a major third and a minor third 
with respect to the root tone.
An inversion of a chord is obtained by transposing the currently lowest tone by an octave.
For example, an inversion of the C major chord C-E-G becomes E-G-C having $\{0, 3, 8\}$ semitones;
yet a further inversion yields G-C-E having $\{0, 5, 9\}$.  
The inversions of a chord do not alter the harmonic effect of it, because a tone sounds the same as its octave.

There are also diminished triad having $\{0, 3, 6\}$ semitones (built from two minor thirds),
and augmented triad having $\{0, 4, 8\}$ semitones (from two major thirds).

If another third is stacked on top of the existing triad, a four-tone seventh chord is obtained having either
$\{0, 4, 7, 10\}$ (as in dominant seventh) or $\{0, 4, 7, 11\}$ (as in major seventh)
or $\{0, 3, 7, 10\}$ (as in minor seventh).
Because there is an interval of second or seventh that creates dissonance, 
all the seventh chords sound unsettling or unstable, bring a feeling of motion toward a nearby stable rest chord.

Further adding thirds successively can yield five-tone ninth chords, six-tone eleventh chords, etc. 
Table 5.1 lists the interval names of tones in a diatonic scale for given numbers of semitones.  
Those above octave can also be considered as a tone within an octave of one octave higher.
For example, the ninth is a second of one octave higher.

\begin{table}
\caption{Intervals related to number of semitones.}
\begin{center}
\begin{tabular*}{0.75\textwidth}{@{\extracolsep{\fill}} lcc}
\hline
\hline
\\ 
 no. of semitones & Interval  & Synonym  \\
\hline
\\
  0 & unison & tonic  \\
  1 & minor second & flat second \\
  2 & major second & second  \\
  3 & minor third & --  \\
  4 & major third & --  \\
  5 & perfect fourth & --  \\
  6 & augmented fourth & diminished or flat fifth  \\
  7 & perfect fifth & --  \\
  8 & minor sixth & augmented or sharp fifth  \\
  9 & major sixth & --  \\
  10 & minor seventh & flat seventh  \\
  11 & major seventh & --  \\
  12 & octave & --  \\
  13 & minor ninth & flat ninth  \\
  14 & major ninth & ninth  \\
\hline
\hline
\end{tabular*}
\end{center}
\end{table}

\section{Harmonic progression} \index{Chord progression}
A harmonic progression (or chord progression) is a goal-directed succession of musical chords, or chord changes that
``aims for a definite goal'' of establishing (or contradicting) a tonality founded on 
a key or tonic chord.\footnote{Tonality is a system
in which specific hierarchical pitch relationships are based on a key ``center'', or tonic.
It is most often used to refer to the major-minor tonality which is also called diatonic tonality.}
It is used to harmonize a composition, as a harmonic simultaneity succession
offering an ongoing shift of level that is essential to the music.
A change of chord, or ``chord change'', generally occurs on an accented beat,
so that chord progressions may contribute significantly to the rhythm, meter and musical form of a piece,
delineating bars, phrases and sections.

The diatonic harmonization of any major scale results in three major triads.
They are based on the first, fourth, and fifth scale degrees, i.e., the tonic, subdominant, and dominant triads.
These three triads include, and therefore can harmonize, every note of that scale.

The same scale also provides three relative minor chords,
one related to each of the three major chords.
These are based upon the sixth, second, and third degrees of the major scale 
and stand in the same relationship to one another as do the three majors,
so they may be viewed as the first, fourth, and fifth degrees of the relative minor key.
Separate from these six common chords the seventh degree of the scale results in a diminished chord.

In classical notation, chords built on the scale are numbered with Roman numerals.
The D major chord will be figured ``I'' in the key of D, but ``IV'' in the key of A.  
Minor chords are signified by lower case Roman, so D minor in the key of C would be written ii.
The names of chords within a key is given in table 5.2, 
with those built on the first, fourth, and fifth degrees of the scale being the major chords (I, IV, and V)
and those on the second, third, and sixth the minor chords (ii, iii, and vi).  
The ``Subtonic'' chord built on the seventh degree of the scale (ti-re-fa) is actually a diminished chord.

\begin{table}
\caption{Names of chords within a key}
\begin{center}
\begin{tabular*}{0.75\textwidth}{@{\extracolsep{\fill}} lcc}
\hline
\hline
\\ 
  Chord symbol & Chord name  & Triad  \\
\hline
\\
  I & Tonic & do-mi-so  \\
  ii & Supertonic & re-fa-la \\
  iii & Mediant & mi-so-ti \\
  IV & Subdominant & fa-la-do  \\
  V & Dominant & so-ti-re  \\
  vi & Submediant & la-do-mi  \\
  vii & Subtonic, or leading tone & ti-re-fa  \\
\hline
\hline
\end{tabular*}
\end{center}
\end{table}

The traditional form of tonal music begins and ends on the tonic of the piece,
and many tonal works move to a closely related key, such as the dominant of the main tonality.
Establishing a tonality is traditionally accomplished through a cadence,
which is two chords in succession yielding a feeling of completion or rest.
The most common cadence is the dominant seventh chord to tonic chord (V7--I) cadence.
Other cadences are considered to be less powerful.

\subsection{Simple chord progressions}
Diatonic scales such as the major and minor scales lend themselves particularly well
to the construction of common chords because they contain a large number of perfect fifths.
Such scales predominate in those regions where harmony is an essential part of music,
as in the classical music.  

Alternation between two chords may be thought of as the most basic chord progression.
Many well-known melodies are built harmonically upon the mere repetition of two chords of 
the same scale such as the tonic I and the dominant V (which sometimes with an added seventh)
or repeated I-IV chords.
The strongest of all progressions involves the root of the chord moving down a fifth, like V-I.
Extending the V-I progression backwards creates a whole series of root motions by fifth down,
i.e., the circle-of-fifths sequence, as
iii-vi-ii-V-I.

Three-chord tunes are more common, since a melody may dwell on any note of the scale.
Often the chords may be selected to fit a pre-conceived melody,
but just as often it is the progression itself that gives rise to the melody.
The three-chord I-IV-V progression can be placed into a four-bar phrase in several ways in popular music.
Many folk songs and other simple tunes can be accompanied using only the I, IV, and V (or V7, dominant seventh)
chords of a key, 
a fact greatly appreciated and often exploited by many beginning guitar players.
However, there are certain rules to follow for the simple chord progressions among these three chords.
The tonic chord I can usually move to any other chords.  Thus, I-V or I-IV progression should be good 
for there are shared tones in both I and V as well as I and IV.
The progression of subdominant IV can be either IV-I or IV-V.  
But the V-IV progression should be avoided (for there is no shared tones in V and IV), 
although the V-I progression produces a strong cadence.

\subsection{Minor chord progressions} \index{Minor scale}
Since minor scales follow a different pattern of intervals from major scales,
they will produce chord progressions with significant differences from major key chord progressions.

The actual chords created using the major scale and its relative (natural) minor scale are the same.
For example, the chords in A minor and those in C major are the same, 
but their usage is different.
If the key is C major, the chord progression will likely make it clear that C is the tonal center of the piece,
e.g., by featuring the bright-sounding (major) tonic, dominant, and subdomiant chords 
(C major, G major or G7, and F major), particularly in strong cadences that end on a C chord.

If the piece is in A minor, on the other hand, it will be more likely to feature 
(particularly in cadences) the tonic, dominant, and subdominant of A minor (A minor, D minor, and E minor chords
as i, iv, and v, with the mediant III, submediant VI, and subtonic VII being major and the supertonic ii being diminished).
These chords are also available in the key of C major, but they will not be given such a prominent place.

The flavor of sound that is produced by a major chord, with a minor seventh add (as in dominant seventh),
renders a particularly dominant (wanting-to-go-to-the-home-chord) sound,
which in turn brings out a stronger feeling of tonality to a piece of music.
Because of this effect, many minor pieces change the v chord to a major chord so that it is a dominant seventh,
even though that requires using a note that is not in the key scale (i.e., by raising the seventh scale note by one half step).
This leads to the usage of a ``harmonic'' minor scale which raises the seventh note of the 
natural minor scale by one half step.
Changes can also be made to the melodic lines of a minor-key piece 
to make it more strongly tonal,
e.g., by raising both the sixth and seventh scale notes by one half step when the melody is ascending 
as in melodic minor scales.

In standard notation, the key signature comes right after the clef symbol on staff.
It may have either some sharp or flat symbols on particular lines or spaces.
The name of a major key is located one half step higher than the last sharp or the second-to-last flat in the key signature.
Since a major scale share the same tones as its relative natural minor scale,
the key signature does not tell whether the music is in a major key or a minor key.
To identify whether a piece of music is in a major or a minor key, it is often effective to
examine the chords at the very end, and at other important cadences such as places where
the music comes to a stopping or resting point.
Most pieces end on the tonic chord.
If the final chord is the tonic of either the major or minor key for that key signature,
the key is certainly determined.

What is amazing here is the fact that when listening to music,
our brains are perceiving the tones as a single entity within a defined pattern,
where how many times particular notes are sounded,
whether they appear in strong or weak beats, 
and how long they last, are kept track of (without our conscious awareness).
By processing such perceptional inputs, the brain makes an inference about the 
major or minor key 
the music is in, despite that the major scale and its relative minor scale use exactly the same
set of pitches.

\subsection{Modulation} \index{Modulation}
Sometimes a piece of music temporarily moves into a new key.
This is called modulation.
Modulation is a very common practice in traditional classical music;
long symphony and concerto movements almost always spend some time in a different key
(usually a closely related key such as the dominant or subdominant,
or the relative minor or relative major), to keep things interesting.
In most styles of music, a slow, gradual modulation to a the new key (and back) 
sounds more natural,
and abrupt modulations can produce unpleasant and jarring feelings.

\section{Melodic harmonization} \index{Melodic harmonization}
To harmonize a single-line music (monophony), 
it might be the simplest by harmonizing a diatonic third above the original melody.
Some people say ``one third fits all.''
The inverse of a third is a sixth, and a sixth is another very nice way to harmonize a melody.

When harmonizing a melody, certain intervals work almost all the time, 
and others are very hard to use.

\begin{itemize}
\item
Unison/octave---Not really a harmony per se, but the effect of doubling a melody can be a nice way to add some textures.

\item
Seconds---They verge on the edge of tension and dissonance and should be handled with care.  They can work at 
certain points in a harmony for adding color, but they are rarely used consistently.

\item
Thirds---They always sound nice, and work great in parallel, too.

\item
Fourths---They can be nice, but not in parallel.  Parallel fourths are one of the main no-no rules of voice leading.
However, since there is a fourth interval from the fifth of a triad to the (inverted) root, 
a fourth can be just the right interval.

\item
Fifths---They are also consonant intervals that work well.  
But they should not be used in parallel, as 
they break the other rule of voice leading. 
When harmonized with straight parallel fifths, the music sounds like a monophony.

\item
Sixths---They are the inverse of thirds, so they always sound good all the time 
especially in parallel motion.

\item
Sevenths---They are another dissonant/tense interval.  They may work at certain points, 
but in general they should not appear in succession one after another because sevenths will not sound consonant.
\end{itemize}

The tense intervals, such as the seconds and sevenths, are not generally something to avoid.
A bit of tension and release is what makes the music sound interesting,
so using those intervals sparingly may add just the perfect color to the music.

\section{Voice leading}  \index{Voice leading}
In musical composition, voice leading is the term for 
a decision-making consideration when arranging voices (or ``parts''),
namely, how each voice should advance from each chord to the next.
The individual melodic lines (called voices) that make up a composition 
interact together to create harmony, where the vertical aspect (chords) 
and the horizontal aspect (voices) are equally important.
In other words, voice leading is the relationship between the successive pitches of 
simultaneously moving voices.
It may be described as parsimonious following ``the law of 
the shortest way'' \cite{schoenberg83},  
so each of the voices should move no more than necessary
taking the smallest possible step or leap.

In music of four parts (e.g., bass, tenor, alto, and soprano),
there are certain general rules of voice leading that should be followed.

\vspace{10 mm}

\noindent
Chord construction (the vertical aspect) rules:
\begin{itemize}
\item
Double the root in any root position chord

\item
Double the soprano in a first inversion chord

\item
Double the bass in a first inversion diminished triad

\item
Double the bass in a second inversion chord

\item
Do not double the leading tone or any altered note

\item
Keep the voices of soprano and alto and the alto and tenor within one octave ; more than
an octave may exist between the tenor and bass voices (the chord spacing rule)
\end{itemize}

\noindent
Voice leading (the horizontal, temporal aspect) rules:
\begin{itemize}
\item
Repeat tones when possible (the common tone rule, i.e., tones common to 
consecutive chords should be retained in the same voices)

\item
Avoid parallel unisons, octaves, and perfect fifths 

\item
Resolve active tones to stable ones

\item
Do not cross voices

\item
Avoid large melodic leaps (the principle of pitch proximity, i.e., 
to maintain the coherence of a single auditory stream by 
having close pitch proximity in successive tones),
with the exception of an octave leap in the bass voice

\item
Strive for a balance of similar, contrary, oblique, and parallel motion between parts
\end{itemize}

\noindent
There can be several directions in melodic motion
for smoothing the melody: \index{Melodic motion}
\begin{itemize}
\item
Similar motion---when two or more voices move in the same direction

\item
Parallel motion---when two or more voices move in the same direction with the same interval 

\item
Contrary motion---when one voice moves in one direction and another voice in the opposite direction

\item
Oblique motion---when one voice stays the same and another voice moves toward or away from it
\end{itemize}

\index{Contrapuntal music}
In traditional contrapuntal music, voice leading is typically derived from the rules of 
counterpoint, which is the relationship between two or more voices
that are harmonically interdependent but independent in pitch contour and rhythm. 
In general, counterpoint involves musical lines that sound very different
and move independently from each other but also sound harmonious when 
played simultaneously.
It focuses primarily on melodic interaction while only secondarily on the ``vertical'' features
of the harmonies.  However, the harmonies are produced by 
such melodic interaction.
It is impossible to have simultaneous musical lines without producing harmony,
and also impossible to have harmonic chord progression without linear activity.
Bach's counterpoint---often considered the most profound synthesis of the two dimensions
ever achieved---is extremely rich harmonically and always clearly directed tonally,
while the individual lines remain fascinating.

\section{Harmony perception}  \index{Harmony perception}
From the psychological point of view, there seems to be grouping principles 
for sound perception.
Grouping is partly an automatic process, referring to that much of it happens rapidly
in our brains and without our conscious awareness.
It may be described as an unconscious process that involves inferencing,
or logical deductions about what objects in the world are likely to go together
based on a number of features or attributes of the objects.
With musical pitches, the rules for what are likely to go together are key and harmony.
A musical key is the tonal context for a piece of music, with which there is a central set
of pitches that the music comes back to, a tonal center, the key.
The key can change during the course of the song (called modulation);
but there is usually a key within certain portion that holds for 
a relatively long period of time during the course of the song.
The attribute of pitch in music functions within a scale or a tonal-harmonic context.
We hear it within the context of a melody and what has come before,
and we hear it within the context of harmony and chords that are accompanying it.\cite{levitin07}

As part of the game of expectation with auditory grouping, 
variations of a theme are often perceived as the same melody despite that 
many notes are actually changed.
Improvisatorial addition of ornaments, especially popular in playing the Baroque music, 
does not change the overall melody.
In classical guitar, there is a special technique called tremolo \index{Tremolo}
that renders one of the most intriguing effects.  
It is usually played as four sixteenth notes in each beat with one bass note 
followed by three repeating melody notes.
When the melody notes are played evenly and fast enough,
an auditory illusion of a sustained note without gaps is created 
even though there is a missing sound of a sixteenth note in each beat as replaced by
the bass note.\cite{parkening97}

In attempting to explain harmony perception, people found that
the first few overtones in a musical sound of any one note add up to its major chord.
Thus, the origin of the major triad and hence its high perceived sonority 
can be explained by overtones.
But overtones are known to fail when used to explain the origin of the minor chord.
This lead to the development of a theory based on the notion of periodicity with
continued fraction expansion and Euclidean algorithm.\cite{stolzenburg09}
This approach shows good correlation to perceived sonority and can be used to 
describe the origin of common musical scales and complex chords.

The perceived sonority of the triads of diatonic music,
such as major $>$ minor $>$ diminished $>$ augmented,
was explained by an analysis of a three-tone ``tension'' factor, based on calculation of the 
relative size of neighboring intervals.\cite{cook09}
It was shown that the relative size of the intervals among the partials in triads 
determines the major and minor modalities of chords.\footnote{A series of 
intervals used to construct a scale is called a (musical) mode, 
which refers to a type of scale such as the ``major'' mode or ``minor'' mode.  
Modality refers to the configuration of intervals between notes in
a scale. \index{Mode, modality}}
It was found that the ``frequency code'' (i.e., rising or falling pitch 
in the intonation of speech and in the vocalizations of animal species) 
is also applicable to the direction of tonal movement.
Each and every pitch rise or fall contains an implication of strength or weakness
among competing animal species:
upward pitch movement implies the negative affect of social weakness, sadness
politeness, assent, questions, whereas 
downward pitch movement implies the positive affect of social strength,
dominance, victory, commands, assertions, happiness.

\newpage
\thispagestyle{empty}


\chapter{Musical Instruments}

A musical sound consists of harmonized sound waves mostly with 
frequencies of integer relationship.
Standing waves on a vibrating object with specifically designed geometry
typically resonate at its natural frequencies of integer relationship,
due to wave reflections at boundaries of sudden changes in 
the acoustic impedance.\cite{rayleigh45} \index{Standing wave}
A well known example is the vibrating string with two fixed ends,
which define the two nodes of 
the fundamental mode (see figure 1.1).\footnote{In a standing wave,
there are points remaining motionless called ``nodes'' 
whereas those having the largest vibration amplitudes are called 
``antinodes''} \index{Nodes, antinodes}
In an open pipe, the vibrating air column has 
standing waves with common antinodes located at the two ends
where the sound waves suddenly lost pipe wall restriction.
The fundamental wave mode has its node located at the 
middle point of the pipe (of a uniform cross section).
In a pipe with one end closed and one open,
the standing waves will have a node at the closed end and 
an antinode at the open end.

Sometimes the sound generated directly from a source is ``noisy''
(or containing irregular overtones) by itself.  
For example, the sounds from the mouthpieces of wind instruments 
alone usually do not contain frequencies of simple integer 
relationship, but with lots of extra frequency components.
When the noisy vibrations of the mouthpiece enter 
the attached instrument with regularly shaped geometry such as a pipe,
only those components from the mouthpiece associated with 
the right wavelengths for the pipe can form the desired 
standing waves resonating in the pipe.
Therefore, the sound transmitted out of the instrument 
becomes the musical sound of a tone.
In this case, the tube in the instrument serves as a sound purifying 
resonator. 

As the source of musical sound, musical instruments 
are usually made of strings with fixed ends
or pipes with at least one open end, because of their natural
capabilities for producing harmonized standing waves.

\section{Chordophones---string instruments}
Chordophones are musical instruments in which standing waves 
that produce sound
are initiated in the strings of the instruments,
such as guitars, violins, harps, pianos, etc.
(some of which are shown in figure 6.1).

\begin{figure}[h] \label{StringInstruments}
\includegraphics[scale=0.64]{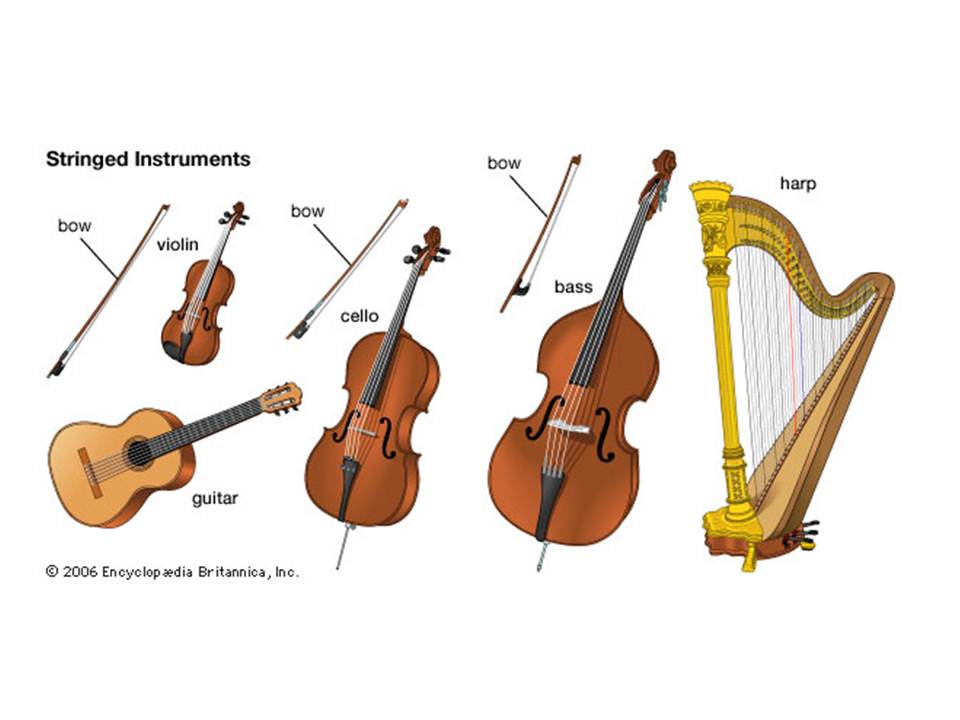}
\caption{Some of the common string instruments.}
\end{figure}

When a string with fixed two ends is plucked (like in guitars and harps) 
or bowed (like in violins) or thumped (like in pianos), 
a transverse standing wave pattern is formed along the string 
with the vibrating velocity in the direction perpendicular to
the string.
As shown in section 1.1 the vibrating string can be decomposed into
a series of sinusoidal components, namely in terms of a Fourier series, 
with angular frequencies given by
\begin{equation} \label{string_vib_freq}
\omega_n = \frac{n \pi}{L} \sqrt{\frac{T}{\mu}} \quad , 
\quad n = 1, 2, 3, ... \, ,
\end{equation}
where $n$ is an integer, $L$ the length of the string, 
$T$ the tension of string, and $\mu$ the linear mass density of string.

Here $\sqrt{T/\mu}$ is the speed of a traveling wave along the string,
which constitutes half of a standing wave.  
According physics of waves, 
each of the wave component of frequency $\omega_n$  
corresponding to a wavelength $L/(2 n)$.
Thus the fundamental mode, as the one with the lowest frequency,
has the longest wavelength equal to twice the string length.
It contains smallest number of nodes, one at each end.
The frequency of the fundamental wave, namely the fundamental frequency
($f_1$ in figure 1.1), defines the pitch of the tone (or note)
that the instrument produces.
Besides the fundamental wave, there are also standing waves 
of frequencies $\omega_n$ with $n = 2, 3, 4, ...$ 
called overtones, which determines the tone color.
The overtone of frequency $\omega_n$ has $n + 1$ nodes and 
$n$ antinodes, more than those of the fundamental wave
with $2$ nodes and only one antinode.

\subsection{Tuning the string instruments}
Because the (open) string lengths---scale lengths---on an instrument 
are usually 
given by the instrument design and the 
values of physical density of strings cannot be altered by 
anybody other than the string manufacturer,
the only thing that can be varied on a string instrument 
is the string tension for adjusting the pitch of a note.  

In the standard tuning of a guitar, the six open strings \index{Guitar tuning}
are tuned to the pitches of E2, A2, D3, G3, B3, and E4.\footnote{The guitar
is a transposing instrument, with its actual pitches sound one octave lower than notated.}
Because guitar has a fretted fingerboard, 
strings can be easily tuned by adjusting the string tension
with the method of comparing 
an open string note with the fretted same note on 
a neighboring string, as described in 
section 2.3 with just one string being set in tune 
(e.g., by a tuning fork).
However, because the frets are placed based on 
equal tempered tuning which is a technical compromise from 
the perfect tuning of the just intonation for a uniformly 
distributed semitone intervals (cf. section 2.1.2),
most of the notes are supposed to be slightly out of tune. 
Therefore, guitarists often fine tune 
their instruments to make the prominent chords in the piece they
are playing sound good, reaching a balance or compromise between them.\cite{parkening97}
A note can also be pushed flat or pulled sharp by pushing or pulling 
parallel to the string with the fretting finger,
which works especially well in higher fret positions 
where intonation is more likely to be a problem.

The four open strings of a violin are tuned \index{Violin}
in perfect fifths to the notes of G3, D4, A4, and E5.
To tune a violin, the A string is first tuned 
by adjusting its tension to a standard
pitch (e.g., 440 Hz), using a tuning device or 
another instrument (e.g., a fixed-pitch instrument such as a piano).
The other strings are then tuned against each other in intervals
of perfect fifths by bowing them in pairs.
Because violins use unfretted fingerboards, 
in theory every note can be
produced perfectly by continuous fine adjustment of the finger position.
Without frets, the player must know exactly where to place the fingers
on the strings against the fingerboard to produce sound with 
good intonation.  
Through practice and ear training,
the violinist's left hand finds the notes intuitively by muscle memory.
For example, there are nine notes in first position,
where a stopped note sounds a unison or octave with another open string,
causing it to resonate sympathetically.
Thus, when unaccompanied, a violinist usually does not play consistently
in either the equal tempered or natural (just) scale,
but tends on the whole to conform with the Pythagorean scale.

A standard piano has 88 keys, each with a set of 3 strings \index{Piano}
(except those bass keys which have a set of 2 strings each) stretched 
separately between the hitch pins and tuning pins on a large soundboard.
Therefore, tuning a piano is a time-consuming task that 
requires lots of patience.
It is the act of making minute adjustments to the tensions of
the strings (by turning the tuning pins)
to properly align the intervals between their tones.
For playing music in different keys, a piano is typically tuned 
according to the equal temperament (see section 2.1.2).
However, fine piano tuning requires an assessment of the 
interaction among notes, which is different for every piano,
thus in practice requiring slightly different pitches from 
any theoretical standard.
Similar to the fretted string instruments like guitars,
sometimes pianists would want
their pianos fine tuned to make the prominent chords in the piece they
are about to play sound particularly good, slightly compromising the tones of 
other keys.

Standard harps have their strings stretched between the 
knots tied on the sounding-board and the tuning pegs on the ``crossbar'' neck 
placed at specific intervals and at specific distances from the soundboard.
Thus, the length of each string is fixed and tuning of which involves adjusting the 
tension of the string through turning the corresponding tuning peg.
Through the mechanical action of seven pedals, each affecting the tuning of 
of strings of one pitch-class through varying the lengths of them 
by pressing them against certain contact points (similar to frets), the pitches of 
the strings can be changed among flat, natural and sharp.
Each pedal has three positions, with the top position corresponding to 
the largest string length for obtaining the flat notes.
Therefore, the harp's native tuning (i.e., the open-string tuning)
 is to the scale of C-flat major.
If all pedals are set in the middle position, the scale of C-major is obtained.
In the bottom position for shortening the string again to create a sharp,
the scale of C-sharp major is obtained by setting all seven pedals in the bottom position.
In each position the pedal can be secured in a notch so that the foot 
does not have to keep holding it in the desired position.
The concert harp typically has six and a half octaves ($6 \times 7 + 1 + 4$$=47$ stings),
with notes ranging from three octaves below middle C-flat to three and a half
octaves above, usually ending on G-sharp.

\subsection{Playing the string instruments}
All string instruments produce sound from 
one or more vibrating strings,
with the sound energy being transferred to the air by
the soundboard of the instrument. 
Common methods to make the strings vibrate are 
plucking, bowing, and striking.

\begin{figure}[h] \label{TriangleWave}
\includegraphics[scale=0.62]{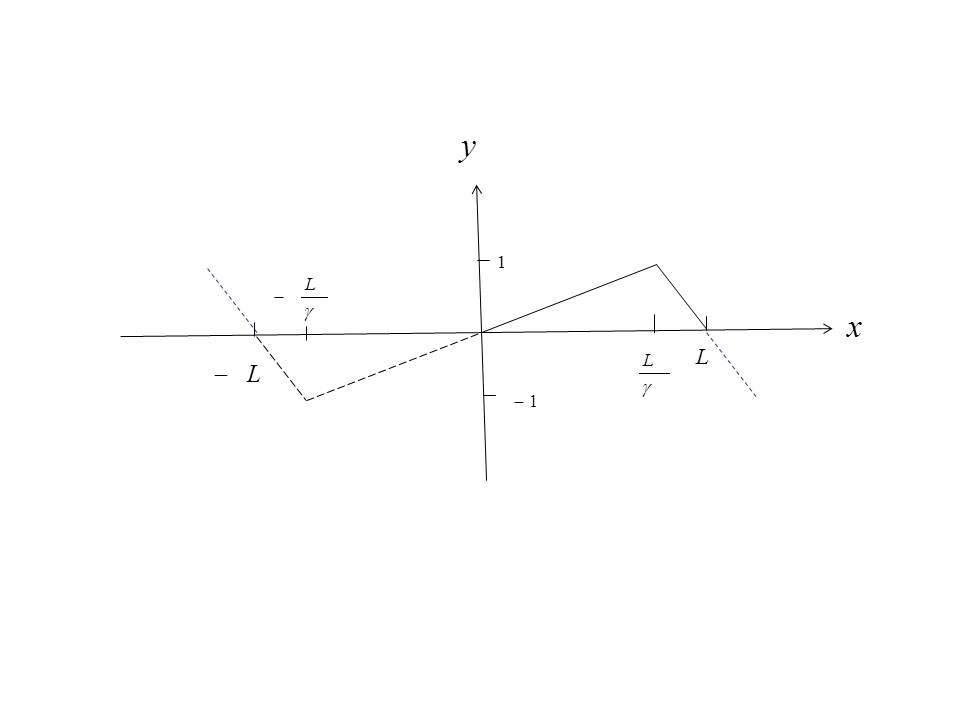}
\caption{A general form of triangle wave with different peak positions 
according to the value of $\gamma$.}
\end{figure}

Plucking is normally used in playing guitar, lute, 
harp, banjo, mandolin, either with a finger or by 
some type of plectrum.
It is also occasionally used in instruments normally played by bowing.
When a finger plucks a string, it effectively deforms 
the string away from its straight equilibrium shape 
into a ``tent'' shape or a triangle wave with the maximum displacement occurs at 
the plucking point or plucking position.  
This is mathematically considered as the ``initial condition'' 
for solving the partial differential wave equation (1.1),
which determines those (Fourier) coefficients in a general solution
in terms of the Fourier series. \index{Fourier series}
For a general triangle wave shown in figure 6.2, we have
\begin{eqnarray} \label{TriangleWaveEq}
f_{\gamma}(x) = \left\{
\begin{array}{lll}
 -1-\frac{\gamma}{(\gamma-1) L}\left(x+\frac{L}{\gamma}\right)  \quad , 
\quad -L \le x \le - \frac{L}{\gamma} \\ \quad \quad   
\frac{\gamma}{L} x \quad \quad \quad \quad , \quad \quad
-\frac{L}{\gamma} \le x \le \frac{L}{\gamma} \\
 1-\frac{\gamma}{(\gamma-1) L}\left(x-\frac{L}{\gamma}\right)  \quad , 
\quad \frac{L}{\gamma} \le x \le L \\
\end{array} \right.  \, ,
\end{eqnarray}
which can be written in terms of a Fourier series
\begin{equation} \label{Fourier_TriangleWave}
f_{\gamma}(x) = \sum_{n=1}^{\infty}b_n \sin\left(\frac{n \pi}{L} x\right)  \quad  
\, ,
\end{equation}
with the Fourier coefficient
\begin{equation} \label{bn}
b_n = \frac1L \int_{-L}^L f_{\gamma}(x)  \sin\left(\frac{n \pi}{L} x\right) dx 
= -\frac{2 \gamma^2 (-1)^n}{n^2 \pi^2 (\gamma - 1)} 
\sin\left[\frac{n (\gamma - 1) \pi}{\gamma}\right] \quad  .
\end{equation}
Those Fourier coefficients define the frequency spectrum,
which can be varied by changing the plucking position along the string.
For a string with two fixed ends at $x = 0$ and $L$, 
we obtain the typical triangle wave that peaks at $L/2$ (i.e. $\gamma = 2$) with 
\begin{equation} \label{bn_triangle}
b_n 
= -\frac{8 (-1)^n}{n^2 \pi^2} 
\sin\left(\frac{n \pi}{2}\right) \quad  ,
\end{equation}
if the plucking position is at the middle $x = L/2$.  
When the plucking position is moved to somewhere near one of the ends,
e.g., approaching $x = L$,  the waveform becomes a sawtooth shape at $\gamma \to 1$,
which has the Fourier coefficient
\begin{equation} \label{bn_sawtooth}
b_n =  
 - \lim_{\gamma \to 1} \frac{2 \gamma^2 (-1)^n}{n^2 \pi^2 (\gamma - 1)} 
\sin\left[\frac{n (\gamma - 1) \pi}{\gamma}\right] 
= -\frac{2 (-1)^n}{n \pi} \quad  .
\end{equation}
It becomes clear that the triangle wave with peak at $L/2$ has fewer harmonics,
namely there are only harmonics with $n$ as odd number, and the higher harmonics
have amplitude decreases with $n$ as $1/n^2$,
where as the sawtooth wave has more harmonics with the higher mode
amplitude decreases at a much slower rate with $n$ as $1/n$.
This is why
a common technique used by guitarists for varying the tone color 
is actively changing the plucking position, 
moving toward the bridge for the ``hard'', brittle, metallic sound
({\it ponticello}) and over the sound hole (somewhat toward the 
middle between the two fixed points set by the fret and bridge)
for the sweet warm tone ({\it dolce}) \cite{parkening97}
which typically contains very few 
harmonics of lower modes (of small $n$). 

Bowing is a method typically used in the violin family 
(such as violin, viola, cello, and double bass).
The bow consists of a stick with many many hairs stretched 
between its ends.  Bowing the instrument's string
causes the string vibration by a stick-slip mechanism. 
The violin produces louder notes with greater bow speed 
or more weight on the string.  The two methods are not equivalent,
because they produce different timbres;
pressing down on the string tends to produce a harsher, more intense
sound.  A louder sound can also be achieved by placing
the bow closer to the bridge.
The point where the bow intersects the string also influences timbre.
Playing with the bow close to the bridge ({\it sul ponticello})
yields a more intense sound emphasizing the higher harmonics,
and with the bow over the end of the fingerboard ({\it sul tasto}) 
produces a delicate, ethereal sound, 
emphasizing the fundamental frequency.

Striking is yet another method for sound production on a string.
Pianos have steel strings mounted on a large soundboard that
are set to vibrate by being struck with hammers which are controlled by
pressing the keys on the keyboard.\footnote{Pressing a key
on the piano's keyboard causes a felt-covered hammer to
strike steel strings, which are allowed to continue vibrating
at their resonant frequency after the hammers rebound}
Because the striking point on the strings for each key is fixed on a piano,
a pianist cannot technically change the timbre of a tone as easy as a guitarist or violinist
by changing the Fourier coefficients for the overtones.
Guitarists sometimes strike the strings with the technique called 
strumming.
Violin family string instrument players are 
occasionally instructed to strike the string with 
the side of the bow, a technique called {\it col legno}.
This yields a percussive sound along with the pitch of the note.

\section{Aerophones---wind instruments} \index{Wind instruments}
Aerophones are musical instruments in which standing waves
are produced in the air column within the instruments.
Wind, brass, and vocal instruments are all similar in that they all use
standing waves of a vibrating air column in a pipe to generate sound.
Figure 6.3 shows some of the wind instruments.

\begin{figure}[h] \label{WindInstruments}
\includegraphics[scale=0.64]{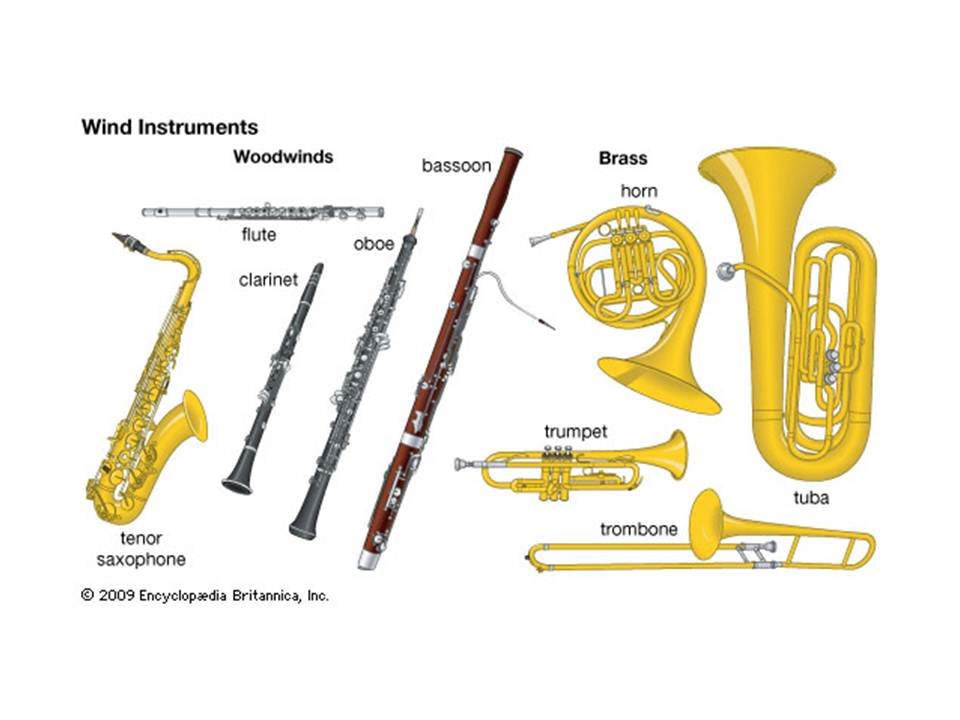}
\caption{Some of the common wind instruments.}
\end{figure}

To create the air column vibration, 
the player directs a stream of air into the instrument.
This air stream is interrupted and chopped into airbursts at a frequency 
within the audible range with some types of reeds (such as
a mechanical reed, a ``lip reed'', or an air reed).
Different tones are created by shortening or lengthening the pipes.
In wind instruments this is commonly accomplished by opening and closing holes on the pipe.
In brass instruments, valves and slides are employed.
For a pipe with both ends open (like a flute), the sound produced contains wavelengths  $2 L/n$
(where $n \ge 1$ and $L$ denotes the length of the pipe) due to the air column resonances.
If a pipe has one end closed (like a clarinet), the resonant wavelengths will be $4 L/(2 n - 1)$ (where $n \ge 1$), indicating that only odd harmonics exist for a pipe with one end closed.
Thus the fundamental frequency of the sound from a pipe is determined by the effective length of the air column $L$,
which can be changed by the player with opening some of the tone holes along the pipe.

The harmonics of wind instruments are little more complicated than that of 
string instruments,
because wind instruments have shapes (of more than one dimension)
ranging from cylindrical to conical and the boundary condition at the pipe open end
is often not exactly well-defined.
The open end of a pipe does not coincide exactly with
an antinode of the standing waves, because there is a region of complicated 
geometry for transition from 
the plane wave inside the pipe to the spherical wave outside.
The acoustic impedance does not abruptly drop to zero at the pipe's open end,
but transition to a finite value defined by the radiation impedance.
With an end correction, a more accurate equation for 
angular frequencies in a cylindrical pipe with both ends open
is given by \cite{kinsler82}
\begin{equation} \label{open_pipe_wavelength}
\omega_n = \frac{n \pi \, c}{L + 0.3 d}  \quad , 
\quad n = 1, 2, 3, ... \, ,
\end{equation}
where $d$ is the diameter of the pipe and $c$ the speed of sound in air
(as discussed in section 1.2).  
Similarly, for a cylindrical pipe with one end closed, a more accurate equation
for angular frequencies is \cite{kinsler82}
\begin{equation} \label{close_pipe_wavelength}
\omega_n = \frac{(2n - 1) \pi \, c}{2 (L + 0.4 d)}  \quad , 
\quad n = 1, 2, 3, ... \, .
\end{equation}
If the instrument has a pipe with $L >> d$,
the longest wavelength corresponding to the lowest frequency of a 
wind instrument is basically determined by the physical length of it: $2 L$ for 
flutes (with pipes of both ends open) and $4 L$ for clarinets (with pipes of one end closed).
Both flutes and clarinets have typical length of about $0.6$ m and internal diameter of 
$15$ mm.  
Thus, flutes have the longest wavelength about $1.2$ m and clarinets about $2.4$ m.
As discussed in section 1.2 with the speed of sound in air about $340$ m$/$s,
we have the lowest frequency for a flute about 283 Hz (middle C sharp) and that for a clarinet 
about 142 Hz (one octave below middle C sharp). 

In most wind instruments, the vibration that the player
makes at the mouthpiece is picked up with 
its harmonic components selectively reinforced
to produce a pleasant timbre by the air column inside 
a tube-shaped body.

The human voice is an instrument in its own right.
A singer generates sounds when airflow from the ungs sets the vocal cords (or folds)
into vibration.  The fundamental frequency here (i.e., the pitch) is controlled by 
the length and tension of the vocal cords and the tone color by the 
formation of the vocal tract.
Women generally have shorter, thinner vocal chords than men,
so women's voices have higher pitch.

\subsection{Tuning the wind instruments}
Wind instruments differ significantly from string instruments
in as much as the wind player is solely responsible for the tuning.
In many cases, the player must `make' the note, and get if right every time.
In order to do so, the player needs both a degree of skill with regard to 
blowing technique and an ability to hear and recognize relative differences in pitch.

The fact is that very few, if any, wind instruments play truly in tune.
We know from section 2.1.2 on equal temperament that the common scale tuning 
is a compromise of pitches.
The wind instruments are yet another exercise in compromise.
When two instruments are put together, there is bound to be the odd anomaly floating about.
The job of the player is to even out those `imperfections'.
It is even more common to come across an instrument that simply just does not 
`gel' with a player's personal technique.

\subsection{Playing the wind instruments}
A wind instrument contains some type of resonator 
(usually a pipe), in which a column of air is set into vibration
by the player blowing into (or over) a mouthpiece attached at 
the end of the resonator.
The pitch of the vibration is determined by the length of the 
pipe and by manual modifications of the effective length of 
the vibrating air column. 
There are several methods for obtaining different notes 
with wind instruments.

\begin{itemize}
\item
Changing the length of the vibrating air column 
through opening or closing holes in the side of the pipe,
as can be done by covering the holes with fingers or pressing a key 
which then closes the hole.
An open hole on a tube, if large enough, defines the virtual end of 
the pipe.
This method is used in nearly all woodwind instruments

\item
Changing the length of the tube through engaging valves 
which route the air through additional tubing,
thereby increasing overall tube length for lowering 
the fundamental pitch.
This method is used on nearly all brass instruments.

\item
Lengthening and/or shortening the pipe using a sliding mechanism.
This method is used on the trombone and the slide whistle.

\item
Making the air column vibrate at desired harmonics 
without changing the physical length of the column of air. 
\end{itemize}

\noindent
All wind instruments use a combination of the above methods 
to extend their register, in order to play all notes 
in the diatonic or chromatic scale.

The brass instruments like trumpet, trombone, and French horn
all have closed pipes with long cylindrical sections and 
should therefore only be able to produce odd harmonics.
However, the mouthpiece and the bell can have a significant effect
on the resonant frequencies.  
Without a bell or mouthpiece, the cylindrical piece of pipe 
will reflect all of its standing wave modes at the same point---the
end of the pipe.
But adding a bell to the pipe enables the reflection at 
different points.
The lower the frequency of the mode, the easier it is influenced 
by the flare of the bell.
Thus the lower frequency modes become shorter in wavelength as 
a result of the bell.
This shorter wavelength increases the frequencies of the lower modes.
On the other hand, the mouthpiece also has an effect.
It has a length of approximately 10 cm and a corresponding
fundamental frequency of its own about 850 Hz. 
This frequency is known as the popping frequency because of 
sound that ``pops'' from the mouthpiece 
if it is removed from the instrument and stuck against the hand.
Thus, the mouthpiece retains some of its 
identity even when inserted into the instrument.
Its presence affects the frequencies of the higher modes,
decreasing their frequencies and increasing their prominence 
in the total spectrum of sound.
Together, the bell and mouthpiece cause the sound production of
the brass instruments to behave like an open pipe,
having all harmonics instead of just the odd ones.
The presence of these modified resonance modes provides 
greater feedback to the player and enhances his or her ability 
to find a particular mode. 

The woodwinds are so named because originally they were mostly
constructed from wood or bamboo, although the metal is used in
constructing modern flutes and saxophones 
and plastic is used to make recorders.
To change the pitch of the woodwind, tone holes along the side of the pipe 
are covered and uncovered when playing.
The function of a tone hole is that, when open, it defines a new end of 
barrel of the instrument.
So, a single pipe can actually be tuned into eight different acoustic pipes 
by drilling seven holes along the side of it.
However, choosing the position of a hole, as well as its size, is not as 
trivial as calculating the length of a pipe for a particular frequency.
Theoretically, the sound wave would reflect at an open hole as if the hole 
is the true end of pipe if the size of the hole were as large as 
the diameter of the pipe.\footnote{The effective
acoustic length actually depends on the relative size of the hole 
compared to the pipe diameter; it will be closer to that defined by the hole position
for larger hole size.}
Structurally, however, it is unreasonable to drill the holes as large as the pipe diameter.
On the other hand, if the bore of the instrument were larger than the fingers,
drilling large holes would require other engineering solutions to enable fully plugging 
the hole.
More complications also arise from the small extra volume under the closed hole,
due to the thickness of the pipe. 
It causes the pipe to appear acoustically longer than its actual length.
The amount of sound wave reflection from a hole on a woodwind instrument 
is also frequency dependent, it often decreases with increasing the frequency.
This fact gives the woodwinds their characteristic timbre.
Thus, the position and size of tone holes on woodwinds are not determined 
by equations for frequencies of standing waves in pipes;
the actual woodwind construction is still based on historic rules of thumb 
and lots of trial and error.

\subsection{Vocal instrument} \index{Vocal instrument}
Music of the vocal instrument is performed by one or more singers, 
typically featuring sung words called lyrics.
A short piece of vocal music with lyrics is broadly terms a song.
Vocal music is probably the oldest form of music; it does not require any instrument
besides the human voice. 
All musical cultures have some form of vocal music.

Voicing, or the phonatory process, occurs when air is expelled from the lungs
through the glottis (the space between the vocal folds), 
creating a pressure drop across the larynx (also commonly called the voice box 
or vocal tract which houses the vocal folds). 
When this pressure drop becomes sufficiently large,
the vocal folds start to vibrate.
The minimum pressure drop required to achieve phonation is called 
the phonation threshold pressure;
it is approximately 200-300 N/m$^2$ for humans with normal vocal folds.
The motion of vibrating vocal folds is mostly lateral 
with almost no motion along the length of the vocal folds.
Such lateral motion of vibrating vocal folds serves to modulate the
pressure in the airflow through the larynx; 
this modulated airflow is the main component of the sound 
of most voiced phones.
Like other wind instruments,
the sound that the larynx produces is a harmonic series, consisting of a fundamental tone
accompanies by harmonic overtones.

The vocal folds will not vibrate if they are not sufficiently close to one another, 
or not under the appropriate amount of tension, or not having sufficient pressure drop
across the larynx.
When they vibrate, the vocal folds are capable of producing several different vibratory patterns.
Each of these vibratory patterns appears within a particular range of pitches
and produces certain characteristic sounds.
A particular vibratory pattern of the vocal folds producing a particular series of tones 
of a common quality in human voice is called a vocal register.
Vocal registers arise from different vibratory patterns in the vocal folds.
There are at least four distinct vibratory forms 
that the vocal folds are capable of producing,
although not all persons can produce all of them.
The first of these vibratory forms is known as natural or normal or modal voice,
which is the natural disposition or manner of action of the vocal folds.
The other three forms are known as vocal fry, falsetto, and whistle.
Arranged by the pitch areas covered,
vocal fry is the lowest register, modal voice is next,
then falsetto, and finally the whistle register.

\begin{itemize}
\item
Vocal fry register---it is the lowest vocal register produced through
a loose glottal closure which permits air to bubble through with a popping or 
rattling sound of a very low frequency.

\item
Modal voice register---it is the usual register for speaking and singing,
with the vocal folds being lengthened, tension increasing, and edges becoming thinner
as pitch rises in this register.
A well-trained singer can phonate two octaves or more in the modal register with
consistent beauty of tone, dynamic variety, and vocal freedom,
which is enabled when the singer avoids static laryngeal adjustments 
and allows the progression from the bottom to top of the register to be
a carefully graduated continuum of readjustments.

\item
Falsetto register---it lies above the modal voice register and overlaps the modal register 
by approximately one octave.  
The characteristic sound of falsetto is inherently breathy and flute-like with 
few overtones present.
The essential difference between the modal and falsetto registers lies in the amount
and type of vocal fold involvement.
The falsetto voice is produced by the vibration of the ligamentous edges of the vocal folds,
in whole or in part, and the main body of the fold is more or less relaxed.
It is more limited in dynamic variation and tone quality than the modal voice.

\item
Whistle register---it is the highest register of the human voice.  
The whistle register is so called because the shrill timber of the notes that are produced from
this register are similar to that of a whistle or the upper notes of a flute.
With proper vocal training, it is possible for most women to develop this part of voice.
Children can also phonate in the whistle register and some men can as well 
in very rare instances.
\end{itemize}

In general, singing is an integrated and coordinated act, which is difficult to discuss
in terms of any of the individual technical processes without relating them to the others.
For example, phonation only comes into perspective when connected with respiration;
the articulators affect resonance (e.g., chest voice or head voice); 
the resonators affect the vocal folds;
the vocal folds affect breath control; and so forth.
Singing is not a natural process but is a skill that requires highly developed muscle reflexes.
It does not require much muscle strength but it does need a high degree of 
muscle coordination.
Individuals can develop their voices further through the careful and systematic
practice of both songs and vocal exercises (usually 
under experienced instructors in an intelligent manner).
The purposes for vocal exercises include \cite{mckinney94}
(i) warming up the voice; (ii) extending the vocal range (to its maximum potential);
(iii) ``lining up'' the voice horizontally and vertically;
(iv) acquiring vocal techniques such as legato, staccato, 
control of dynamics, rapid figurations, to comfortably sing wide intervals,
and correcting vocal faults.
Good singers are thinking constantly about the kind of sound they are producing
and the kind of sensations they feel while they are singing.

An important goal of vocal development is to learn to sing to the natural limits of 
one's vocal range without any obvious or distracting changes of quality or technique.
A singer can only achieve this goal when all the physical processes involved in singing
(such as laryngeal action, breath support, resonance adjustment,
and articulatory movement) are effectively working together, i.e., well coordinated.
The first step in coordinating these processes is by establishing good vocal habits in
the most comfortable tessitura (texture or best sounding texture and timbre) 
of the voice before slowly expanding the 
range beyond that.

\subsection{Organs---keyboard aerophones} \index{Organs}
The (pipe) organs consist of many pipes of varying lengths,
some open-ended and some closed at one end, 
most of which are inside the organ case
arranged in a two-dimensional matrix.
Row upon row of pipes of different pitch but of the same ``timbre'' sit on 
compressed air supply channels where pulling a given ``stop'' (in a ``stop-channel'' design)
moves a wooden slat called slider such that the holes in the slider line up with 
the toe holes of pipes
to admit the air into a given channel.
For a given pipe to actually produce sound, 
the valve in its foot must be opened by the player pressing the appropriate key 
on the keyboard.
For example, when an ``eight-foot flute stop'' is pulled and the G4 key is pressed,
all eight-foot flute pipes get compressed air beneath them 
and the valves on all G4 pipes are opened,
but only the G4 eight-foot flute pipe will sound.
The complexity of the organ sound is achieved by the many possible combinations of 
stops on two to five keyboards (called manuals) 
played by the hands and a full pedalboard played by the feet.

As the air enters the pipe through its toe hole on the bottom, 
it hits the sharp edge at the front causing the 
air to vibrate.  
Certain vibration modes that do not have wavelengths fitting exactly in the pipe will
annihilate themselves, while those modes with wavelengths fitting exactly are 
selectively reinforced by resonance
to produce the primary tone and overtone harmonics of the pipe.
As with the typical wind instruments, the length of the pipe determines 
the resonant wavelengths and the corresponding pitch of the tone.
The design and material of the pipe determine the relative intensity of the overtones 
for the timbre of the sound.

The organ's continuous supply of wind allows it to sustain notes for 
as long as the corresponding keys are pressed,
unlike the piano and harpsichord with their sounds decaying 
when the keys are held.
There are also nonlinear effects that will result in a complicated sound 
different from simple linear superposition of the Fourier components.
For example,
the pipe organ sound is strictly harmonic, in spite of the spectrum of 
each pipe is not quite harmonic.
This is because of the phenomenon of cross-talk between different pipes 
due to the nonlinear effect of frequency-locking that makes the sounds from two slightly out of tune pipes 
have a common frequency instead of the expected beating.\cite{abel06}

\section{Idiophones---percussion instruments} \index{Idiophones}

\begin{figure}[h] \label{Idiophones}
\includegraphics[scale=0.72]{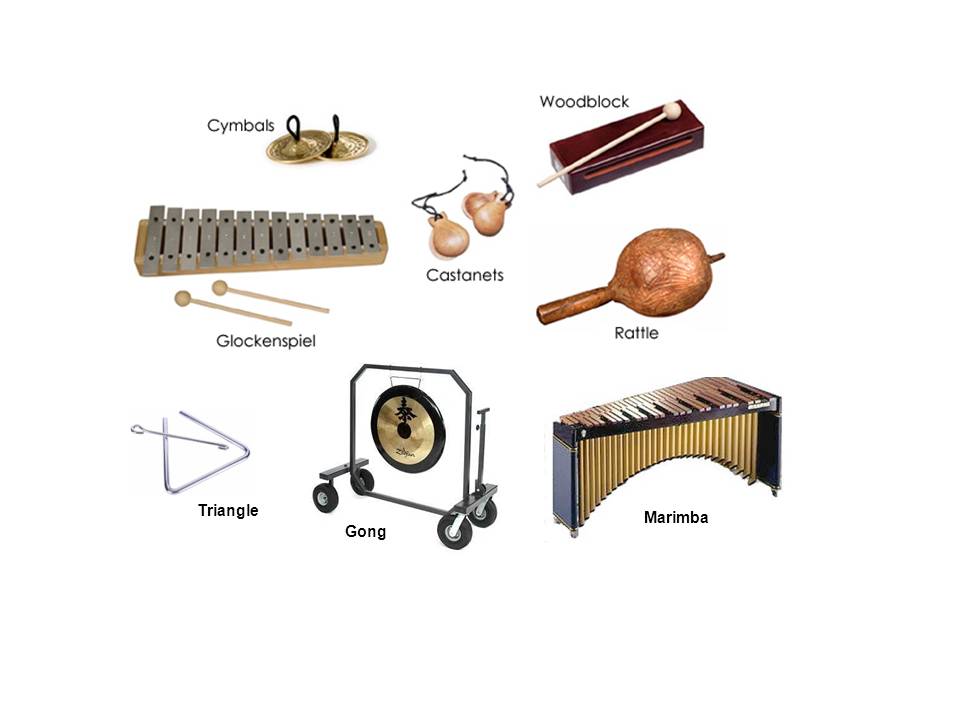}
\caption{Some of the idiophones.}
\end{figure}

\noindent
In idiophones, the vibration of a solid component generates the sound,
such as a bell,
wine glass, xylophone, etc. (as shown in figure 6.4.)
They are made from materials that have their own unique sounds
(such as ceramics, glass, metal, or wood), and are
generally the percussion instruments that are hit,
shaken, moved, or rubbed to produce sound.
Resonators can also be added to these instruments 
to reinforce the sound.
The pitch of the sound is determined by the geometric size and shape
such as the length of a solid bar.
However, the sounds produced when a solid bar or pipe (of uniform cross section) 
is tapped on their sides 
are fundamentally different from that of chordophones and aerophones, because
the frequencies of higher modes here in 
transversely vibrating bars and pipes (of uniform cross section) are 
not harmonic overtones (i.e., not in integer multiples of 
the fundamental frequency).\cite{landau59e, fletcher98}
For example, the characteristic frequencies of a transversely vibrating thin solid rod
with one end ($x = 0$) clamped and the other ($x = L$) free 
are determined by the roots of the equation\cite{landau59e}
\begin{equation} \label{rod_frequencies}
\cos \kappa L \cosh \kappa L + 1 = 0  \quad , \mbox{ with } 
\kappa^4 \equiv \omega^2 \rho S/(E I)  \, ,
\end{equation}
where $\omega$ denotes the angular frequency, 
$\rho S$ the mass per unit length, $E$ Young's modulus, 
and $I$ the moment of inertia (the 
second moment of area of the cross-section), of the transverse vibration of the rod.
The form of (6.9) indicates a complicated relationship among the characteristic frequencies,
unlike the simple integer relationship described by (6.1) for a vibrating string,
and (6.7), (6.8) for a vibrating air column.
Therefore, many percussion instruments do not produce a definite pitch
due to their sounds with a wide range of prominent frequencies that 
are not in harmonic relationship.
These ``unpitched'' or ``untuned'' percussion instruments are played for 
rhythm only.
There are, however, some ``pitched'' or ``tuned'' percussion instruments 
(such as the marimba and xylophone)
can produce an obvious fundamental pitch and can therefore play melody and serve 
harmonic functions in music in addition to the rhythm.

\subsection{Xylophone, glockenspiel---tuned harmonic idiophones} \index{Xylophone}
The xylophone (including marimba) consists of wooden bars that make sound when struck by mallets.
Each bar is an idiophone with varying cross section carved for 
producing a harmonic, clearly defined tone such that it can be
tuned to a pitch of a musical scale.
The bars are usually made of rosewood, padak, or various synthetic materials
such as fiberglass or fiberglass-reinforced plastic which allows a louder sound
It should be played with very hard rubber, polyball, or acrylic mallets.
Concert xylophones typically cover three and half or four octaves.
They also have tube resonators attached below the bars to
enhance the tone and sustain.

When the wooden bars of a xylophone are replaced by metal plates or tubes,
a glockenspiel is constructed which is \index{Glockenspiel}
also composed of a set of tuned harmonic idiophones. 
The glockenspiel is also played with a pair of mallets, like the xylophone.
When struck, the bars produce a very pure, bell-like sound, usually at higher pitch than 
those of xylophones.
The glockenspiel is limited to the upper register and usually covers 
about two and a half to three octaves.
It is a trasposing instrument, with its parts written two octaves below the sounding notes.

\subsection{Tuning fork} \index{Tuning fork}
As a type of idiophone, 
a tuning fork is made in the form of a two-pronged fork with the prongs (tines) 
formed from a U-shaped bar of elastic metal.  
It resonates at a specific pitch when set vibrating by striking it with an object,
and emits a pure musical tone (consisting of a single sinusoidal component as shown in
figure 1.2), after waiting a moment to allow some high overtones to die out.
The particular pitch of a tuning fork generates corresponds to the angular frequency
determined by
\begin{equation} \label{tuning_fork_frequency}
\omega = \frac{1.875^2}{L^2} \sqrt{\frac{E I}{\rho S}}  \quad , 
\end{equation}
where $1.875$ is the smallest positive root of (6.9) for the value of $\kappa L$.
The main reason for the tuning fork to produce a very pure tone, 
with most of the vibrational energy at the fundamental frequency,
is that the frequency of the first overtone is about $6.25$ times 
the fundamental one.\cite{rossing92}
Such an overtone with much higher relative frequency dies out fast after 
the fork is struck, leaving only the fundamental mode ringing.
The reason for using the fork shape is that,
when it vibrates in its fundamental mode, there is a node at the base of each prong.
Thus holding the fork by the handle would not damp the fork vibration significantly.

\subsection{Accordion, harmonica---free-reed instruments} \index{Accordion} \index{Harmonica}
Some people may consider the accordion and harmonica as a wind instruments.
But both accordion and harmonica make all their sounds by means of reeds (which
are thin brass, bronze, or steel tongue-like plates with one end fixed on 
a frame) that are driven to vibrate, much like a swinging door,
by having air flow across them.
The air-driven vibration of free-reed is actually a nonlinear phenomenon of
coupling between the air flow and the reed displacement while both are undergoing 
oscillatory motions, which have inspired quite a few scholarly studies.\cite{cottingham11}
The reed chamber has also been shown to affect both the pitch and the quality of sound.
Therefore, the air-driven free reed vibration is a more involved case than 
a simple harmonic oscillator. 
When driven by an air flow,
each reed vibrates at a frequency near its natural (i.e., plucked) vibration frequency,
which is determined by the mass and stiffness of the reed in 
cooperation with its associated acoustical system.
The vibrating reed tongue chops the air stream that drives its motion,
resulting in complex pressure pulses whose waveform contains abundant higher harmonics 
having frequencies very close to integer ratios.
Only at very high blowing pressures does noticeable departure from harmonicity occur.
In other words, the free-reed vibration is self-excited and does not
require a vibrating air mass in a resonator to transfer a steady air pressure 
into an oscillatory motion.
There are no resonant pipes, as in typical wind instruments, to selectively
reinforce certain harmonic components in free-reed instruments.
Thus the physics of sound production in free-reed instruments 
seems to be more like the harmonic idiophone as the xylophone.\footnote{Nevertheless 
some people also categorize accordion and harmonica 
as the free-reed aerophones}

The accordion has bellows as the most recognizable part of the instrument 
and the primary means of articulation.
The body of accordion consists of two boxes joined together by the bellows.
These boxes house reed chambers for the right- and left-hand manuals, respectively.
Each side has grilles in order to facilitate the transmission of air in and out of the instrument,
and to allow better projection of sound.
It is played by compressing or expanding the bellows while
pressing buttons or keys, causing valves (called pallets) to open for air to flow
across reeds to produce sound.
The instrument is considered a one-man-band as it needs no accompanying instruments.
The player normally plays the melody on keys on the right-hand manual 
and the accompaniment, consisting of bass and pre-set chord buttons, on the 
left-hand manual.

The body of the harmonica is a comb with prongs separating a row of ten or more
chambers.  One plate of reeds (the blow reeds) that are activated by positive pressure 
in the comb (i.e., blowing) is fastened to the top side of the comb with the reeds inside.
Another plate of reeds (the draw reeds) that are activated by negative pressure 
is fastened to the bottom side of the comb with the reeds outside.  
The reed slots (on which the reed overlies in the reed plate) on the blow- and draw-reed
plates are arranged to coincide with each other on opposite sides of the chambers of the comb.
Reeds of the popular ten hole, diatonic harmonica are tuned in a major key 
and arranged in such a way that one can produce the major chord of the key 
by blowing on any three adjacent holes.
Drawing on the first four holes produces the dominant chord of the key.
The harmonica is played by blowing air into it or drawing air out 
with lips placed over individual holes (reed chambers) or multiple holes.
The resulting pressure from blowing or drawing air in the reed chambers causes
a reed or multiple reeds to vibrate for producing sound.
Each chamber has multiple, variable-tuned reeds with one end secured and the other
free to vibrate.
Reeds are tuned to individual tones according to their sizes;
longer reeds produces lower pitch sounds and shorter reeds higher pitch sounds. 
Individual reeds are usually riveted to the reed plate either on the inside 
to respond to blowing or on the outside to respond to suction.

As a standard practice, skilled harmonica players can bend the pitch of 
their instrument by exploiting the coupling of the reeds to the vocal tract of the player
as well as the coupling between the two reeds 
(one for each direction of airflow) sharing a single reed chamber.
Unlike the harmonica, the accordion is blown mechanically and the reeds do not couple
the the vocal tract of the player.
But a player of a standard accordion can still bend the pitch downward with a 
combination of increased pressure from the bellows and restricted airflow
achieved by partially opening a pallet valve.
Such pitch bending is occasionally used for special effect with the accordion;
but it does not provide the kind of flexibility comparable to that
at the disposal of a skilled harmonica player.\cite{cottingham11}

\section{Membranophones---drums} \index{Membranophones} \index{Drums}
Membranophones produce sound with a vibrating membrane over a resonator.
They are musical percussion instruments,
usually consist of a hollow cylinder with a membrane stretched across each end  (see figure 6.5)
which can vibrate when being hit with a stick or hand.
As with Idiophones, some of the membranophones are pitched 
(e.g., with the pitch set by tightening or loosing the tension on the membrane) and some are unpitched. 
Based largely on shape, membranophones are divided into several divisions as follows.

\begin{figure}[h] \label{Drums}
\includegraphics[scale=0.64]{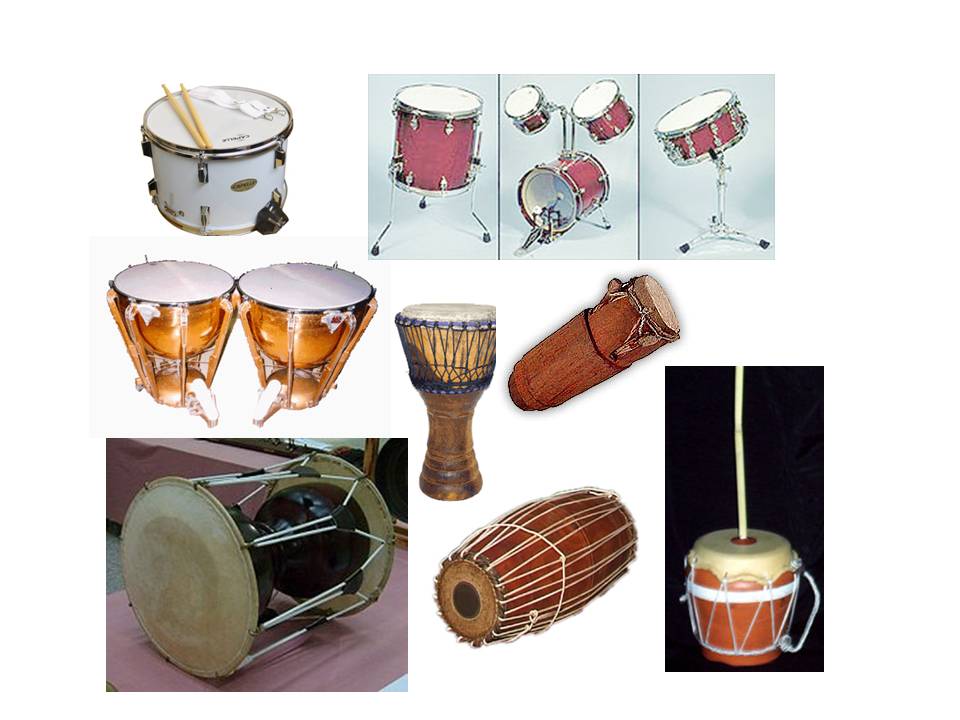}
\caption{Some of the membranophones.}
\end{figure}

\begin{itemize}
\item
Cylindrical drums---straight-sided and generally two-headed, sometimes using a buzzing, percussive string.

\item
Barrel drums---normally one-headed and maybe open at the bottom, with a bulge in the middle.

\item
Hourglass drums (or waisted drums)---hourglass-shaped and generally two-headed, with the drumheads being laced 
onto the body where the laces may be squeezed during performance to alter the drum's pitch.

\item
Glblet drums (or chalice drums)---one-headed and goblet-shaped, and usually open at the bottom.

\item
Footed drums---single-headed and are held above the ground by feet so that the space between the drum and the ground
provides extra resonance.

\item
Long drums---those have length significantly exceed diameter.

\item
Kettle drums (or pot drums or vessel drums)---one-headed, usually tuned to a specific pitch, with a vessel or pot body, and 
are often played in pairs.

\item
Frame drums---those composed of one or more membranes stretched across a frame.

\item
Friction drums---those produce sound through friction by rubbing a hand or object against the drumskin.

\end{itemize}

In addition to drums, there is another kind of membranophone, called the singing membranophone, of which 
the best known type is the kazoo that produce sound by blowing air across a membrane.
This kind of instruments modifies a sound produced by something else, commonly the human voice,
by having a skin vibrate in sympathy with it.

\section{Temperature and humidity effects}
The mechanical (vibrational) properties of all materials change 
in response to a change of temperature
due to the phenomenon of thermal expansion.
For example, the tension of stretched string changes with temperature.
Thus a musical instrument needs tuning whenever its environment changes.

Many musical instruments are either made of wood or have wood components.
Under natural conditions, wood always contains some moisture, which 
helps maintain both its density and strength, among other physical properties.
The density and strength of wood each affect the acoustical and structural properties
of most musical instruments.

Wood swells and shrinks in proportion to the magnitude of change in its moisture content.
Even with properly seasoned wood, the shrinking and swelling can affect the 
acoustics, structure, and more adversely the function of an instrument.
When the structural stability of an instrument is compromised, 
the tuning of it is expected to vary at least.
It is also not uncommon to see cracks develop in the wood of some instruments 
after going through significant changes of temperature and humidity.

The primary source of moisture is the air surrounding the instrument.  
Humidity is the capacity of air to hold moisture.
Absolute humidity is the amount of moisture in a given amount of volume of air.
Relative humidity is the ratio of the absolute humidity to the maximum amount of 
moisture in the same volume of air at a given temperature (i.e., the saturation point).
The capacity of the air to hold moisture is proportional to ambient temperature;
warm air can hold more moisture than cold air.
Thus when the air holding the same amount of moisture is cooled to a lower temperature,
its relative humidity will increase although its absolute humidity is kept constant.
At equilibrium, the moisture content of wood varies directly with the relative humidity
at a given temperature.  On the other hand, the equilibrium moisture content of wood 
varies inversely with temperature if relative humidity is held constant.
The equilibrium moisture content of wood is 
usually more sensitive to small changes in relative humidity
than to small changes in temperature.
Yet the musical instruments are often in the environments that are more likely 
to have a well-controlled temperature than humidity.

Due to slow diffusion of water through the cells of wood, 
the moisture content of wood is slower to reach a state of equilibrium than 
that of air.  
So the instrument response to the temperature and humidity changes is 
not immediate.
It is therefore highly recommended to have the instrument in the performance environment 
for adequate amount of time before fine tuning for the performance.
It is often suggested to keep the musical instruments 
in an environment with temperature at $22^o$ C and relative humidity at 40 percent, 
which is at a reasonable comfort level for humans and is not expected to be too 
far away from the actual performance environment.

\section{Treatment of wood for musical instruments}
One of the most important factors in defining the sound of an instrument is 
the wood used for building it.
Everything else being the same, no two pieces of wood will sound the same.
The quality of the wood include appropriate cutting, careful aging, among other things.
For example, the wood for classical guitar soundboard is typically cut 
by ``quarter sawing''\footnote{The log is cut to blocks of length of about 24 inches
and then cut (or preferably split) into quarters (called billets),
and the boards are then sawn off the resulting flat sides}
such that the grain lines are vertical to the surface.
The wood must also be ``old'', i.e., be naturally aged (preferably by air drying 
for 3 or 4 years).   

Well-constructed musical instruments have had the moisture content of their soundboards
reduced just before installation.
If the moisture content at time of construction is below the lowest equilibrium moisture content
that the instruments will experience in their lifetime, 
no cracks should occur as a result of low humidity.
Of course, the instruments should not be exposed to the air having  too much moisture,
which can cause the soundboard to swell and affect the sound quality and performance.

It has been well-known that many string instruments get better with age.
It is not just the chronological time that does the magic; 
it is the actual playing time as well. 
In fact, the wood that guitars and other instruments are made of 
seems to vibrate more musically the more it is vibrated (namely, played).
In other words, ``use it or lose it.''
The reasons for the improvement by aging may have to do with
subtle changes in the stiffness and flexibility within the cellular structure of the wood,
and the hardening of resins within the cells themselves.
Also, the lacquer, the most common finish on guitars, loses plasticizers and becomes 
more brittle over time.  
As the wood of soundboard ages with vibrations due to frequent playing,
the areas that move a lot in the nodal patterns loosen up and continually flexed 
while other areas of little movements get stiffer.  
Those vibrational patterns are thus ``set'' or ``memorized'' by the well aged soundboard 
to lead to improved musical sounds.
Therefore vintage instruments are so 
revered by players and collectors.

\newpage
\thispagestyle{empty}


\chapter{Music Performance and Practice}
Successful music performance is nothing short of practice.
It is often said ``practice makes perfect'', which seems to be true 
for almost any extraordinary performances, whether they be music, sports, or even stealing 
the British crown jewels without being caught.
Studies have suggested that practice is the cause of hight-degree of accomplishment,
not the so-called ``talent''.
Furthermore, it seems that ten thousand hours of practice is generally required to 
achieve the level of mastery associated with being a world-class ``expert'' 
in anything.\footnote{Ten thousand hours is equivalent to roughly three hours a day,
or twenty hours a week, of practice over ten years}
Of course, this does not address why some people do not get anywhere when 
they keep on practicing, and why some people get more out of their practice 
than others.
Actually to emphasize the correct methods or the quality of practice, 
some people would say ``practice does not make perfect;
perfect practice makes perfect.''  

Nevertheless, the ``ten-thousand-hours'' theory is consistent with 
the way our brains learn.
Learning requires the assimilation and consolidation of information in neural tissue.   
The more repetitive experiences we have with something, 
the stronger the memory-learning trace for that experience becomes.
In other words, increased practice leads to a greater number of neural traces,
which can combine to create a stronger memory representation.
In the neuro-anatomy theory,
the strength of a memory is related to how many times the original stimulus has been 
experiences.
It seems that it takes the brain this (ten thousand hours) long 
to assimilate all that it needs to know to achieve true mastery.
As a matter of fact, no one has yet found a case 
in which true world-class expertise was accomplished in less practice time.\cite{levitin07}

Important as it might be, the amount of practice time is only one factor that is necessary 
for achievement.  The quality of practice is yet another important factor.  
Memory strength also depends on how much we care about the experience.
Caring leads to attention and measurable neurochemical changes.
Thus, the ``perfect practice'' actually involves 
total concentration, all of the body, mind and soul.

\section{To practice correctly}
In the terms of neurobiology, to learn is to establish a new set of neural circuits
and thus a memory trace.
When we perceive something, a particular pattern of neurons 
fire in a particular way for a particular stimulus.
In other words, the act of perceiving entails that an interconnected set of neurons 
becomes activated in a particular way,
giving rise to our mental representation of what is out there in the world.
Remembering can simply be the process of recruiting that same group of neurons
we used during perception to help us form a mental image during recollection.
We re-member the neurons,
pulling them together again from their disparate locations to 
become members of the original club of neurons 
that were active during perception.
Hence, repetitive practice can help strengthen our memory for what to be learned.

To perform a piece of music, we need to learn it by practice until we can
play the correct notes in a fluent manner consistently.
To consistently play the correct notes requires consistent reinforcing our 
memory correctly by practicing the correct notes.
This means that we should try to avoid playing the wrong notes, which 
could make our brains learn the wrong pattern and establish the wrong memory.
Some music teachers would say that it is better not to practice at all than 
to practice incorrectly, for not to reinforcing the wrong notes.

Therefore, it is recommended that when practicing, you should start by 
playing slowly and surely before increasing the tempo.
If there is something of which you are not sure, simply stop and analyze 
to let your brain learn the correct pattern of notes.
For a difficult section or passage, you should take it at a even slower speed 
for playing all the notes correctly at all times.

The practice may be viewed as a simple game of attention,
i.e., what you pay attention to will grow in your reality.
If you react negatively to your mistakes, you are giving the mistakes more power
because more attention is given to them.
The mistakes should be noticed, but you should not react to them to avoid feeding them 
more power.
When practice, you should keep focusing your attention on what you want to grow,
instead of what you try to avoid. 
You should alway bring your mind to playing and listening,
without letting any mistake distract you.
Then all else will fall into place with time.

\section{To play without trying}
It is known that almost all supreme technical display and 
fluent rendition of a music piece come out of the performer who 
simply lets go of the need to control the techniques.
This is a technique some people call ``to do all by doing nothing.''
The paradoxial fact is that the more we try to control, the less control we have.

Wanting control over any situation is usually triggered by the fear of 
losing control over it.  
Tension is a physical manifestation of fear.
Tension makes movement difficult, which then incites more fear in the player.
In other words, more fear equals more tension.  
It is then a never ending cycle.
When we relax our mindset, and stop fearing our loss over control,
the body relaxes and control begins to happen naturally.
This is how we should ``play'' the music.
Great playing is similar to deep meditation,
when the mind goes quiet and observation of the self deepens.
The mind should not be busy making judgments about the playing,
or worrying about what other people would think of the playing,
but should simply be observing the movements of hands peacefully.

\section{The art of musical performance}
In a high quality performance, there is the focus on context,
the overall mindfulness of the music making, and some degree of musical
risk-taking.  
The musical intelligence and technical skills of the performer are used
to help nurture all of the musical details in the work.
A wide range of articulations and dynamics levels are considered,
creating a context-specific musical rhetoric, with attention paid to the 
micro and macro architecture of the music. 
Yet musical performance also takes creativity of the performer,
because the music notations on score sheet cannot tell everything precisely.
Great virtuoso Spanish guitarist Andr\'es Segovia always played his music with 
some delays or onward rushes, against the precise rhythm per music score annotations.
This made his strong personal style legendary.
He once said that it is in this lack of respect (for the annotated rhythm)
that you may define the good artist and the bad artist.
When the performer is fully engaged, 
the music is performed with confidence and interpretive conviction.
It becomes the projection of a cogent realization of the work to the listener.

Some of the greatest performances can hold the listener's ear captive,
having the listener emotionally touched and intellectually stimulated.
Often, the synergy between performer and listener is so complete that 
both are ``in flow'', each completely unaware of ``the self'' and 
the passing of time.
The performer is completely invested in every aspect of the music using 
varied, context-specific tone and color, demonstrating a high sensitivity
to harmony, structure, and line, and manifesting deep comprehension of 
the music through assured projection of its essence.
A performer hoping to one day achieve the highest level of performance 
must first commit to spending all of his or her time, 
in practice as well as on-stage, at the caring and convincing levels.
To reach the magical outcome, the dues must be paid daily.

\section{Success versus excellence}
Success and excellence are often competing ideals.
Being successful does not necessarily mean you will be excellent,
and being excellent does not necessarily mean you will be successful.
Success is attaining or achieving cultural goals, which elevates one's importance
in the society in which he lives.
Excellence is the pursuit of quality in one's work and effort, 
whether the culture recognizes it or not.

Success seeks status, power, prestige, wealth, and privilege.
Excellence is internal--seeking satisfaction in having done your best.
Success is external--how you have done in comparison to others.
Excellence is how you have done in relation to your own potential.
For certain people, success seeks to please men,
but excellence seeks to please God.

Success grants its rewards to a few, but is the dream of the multitudes.
Excellence is available to all, but is accepted only by a few.
Success engenders a fantasy and a compulsive groping for the pot of gold at the end
of the rainbow. 
Excellence brings us down to reality with a deep gratitude for the promise of joy 
when we do our best.
Excellence cultivates principles, character, and integrity.
Success may be cheap, and you can take shortcuts to get there.
You will pay the full price for excellence; it is never discounted.
Excellence will always cost you everything, but it is the most lasting and 
rewarding ideal.\cite{parkening97}

\newpage
\thispagestyle{empty}

\bibliographystyle{plain}

\cleardoublepage
  \addcontentsline{toc}{chapter}{Index}
\printindex

\end{document}